\documentclass[aps,prb,amsmath,amssymb,superscriptaddress,floatfix,showpacs,twocolumn]{revtex4}

\usepackage[pdftex]{graphicx} 
\usepackage{epstopdf} 
\usepackage{subfigure}
\usepackage[usenames,dvipsnames]{color}
\usepackage{bm}

\usepackage{hyperref}

\begin{document}

\title{Coulombic Charge Ice}
\author{P. A. McClarty}
\address{Max Planck Institute for the Physics of Complex Systems, N{\"o}thnitzer Str. 38,01187 Dresden, Germany}
\address{ISIS Facility, Rutherford Appleton Laboratory, Chilton, Didcot, Oxon. OX11 OQX, United Kingdom}
\author{A. O'Brien}
\address{School of Physics, The University of Sydney, Sydney, New South Wales 2006, Australia}
\address{Max Planck Institute for the Physics of Complex Systems, N{\"o}thnitzer Str. 38,01187 Dresden, Germany}
\author{F. Pollmann}
\address{Max Planck Institute for the Physics of Complex Systems, N{\"o}thnitzer Str. 38,01187 Dresden, Germany}	
%\date{} % delete this line to display the current date

\begin{abstract}
We consider a classical model of charges $\pm q$ on a pyrochlore lattice in the presence of long range Coulomb interactions. This model first appeared in the early literature on charge order in magnetite [\onlinecite{Anderson56}]. In the limit where the  interactions become short-ranged, the model has a ground state with an extensive entropy and dipolar charge-charge correlations. When long range interactions are introduced, the exact degeneracy is broken. We study the thermodynamics of the model and show the presence of a correlated charge liquid within a temperature window in which the physics is well described as a liquid of screened charged defects. The structure factor in this phase, which has smeared pinch points at the reciprocal lattice points, may be used to detect charge ice experimentally. In addition, the model exhibits fractionally charged excitations  $\pm q/2$ which are shown to interact via a $1/r$ potential. At lower temperatures, the model exhibits a transition to a long-range ordered phase. We are able to treat the Coulombic charge ice model and the dipolar spin ice model on an equal footing by mapping both to a constrained charge model on the diamond lattice. We find that states of the two ice models are related by a staggering field which is reflected in the energetics of these two models. From this perspective we can understand the origin of the spin ice and charge ice ground states as coming from a dipolar model on a diamond lattice. We study the properties of charge ice in an external electric field finding that the correlated liquid is robust to the presence of a field in contrast to the case of spin ice in a magnetic field. Finally, we comment on the transport properties of Coulombic charge ice in the correlated liquid phase.
\end{abstract}

\maketitle

\section{Introduction}

There has been a great deal of recent interest in the effects of geometrical frustration in condensed matter systems. A significant part of this interest has centered around and been inspired by the discovery and characterisation of the spin ice materials Dy$_{2}$Ti$_{2}$O$_{7}$ and Ho$_{2}$Ti$_{2}$O$_{7}$.  \cite{Bramwell01} These materials were found, in the late nineties, to harbour an extensive residual entropy at low temperatures equal to the entropy of water ice.  \cite{Bramwell01} This experiment was motivated by the observation that the low-lying proton configurations in water ice map onto the configurations of a frustrated Ising model on a pyrochlore lattice. The discovery of residual entropy in these rare earth pyrochlores has led to a flurry of theoretical and experimental works over the last decade that have brought to light a number of other remarkable features of the spin ices. These include their distinctive anisotropic spin correlations controlled by the emergence of a divergence-free constraint analogous to the vacuum Gauss' law in electromagnetism \cite{Henley2011} and the existence of fractionalized excitations that behave like charged particles albeit in a magnetic insulator. \cite{Castelnovo2008}

One of the marvellous features of the spin ice materials is that much of their phenomenology is captured quantitatively by the dipolar spin ice model. The aforementioned properties of the spin ices at low temperatures are shared by a number of other models \cite{Henley2011,Chern2011,Banks2012} and collectively constitute the defining features of so-called Coulomb phases. One such model, which we call the Coulombic charge ice model, was considered a long time ago by Verwey and co-workers \cite{Verwey1947} and by Anderson \cite{Anderson56} as a potential means to understand the charge ordering in magnetite, Fe$_{3}$O$_{4}$, an iron oxide spinel with equal numbers of Fe$^{2+}$ and Fe$^{3+}$ ions on the B sites. \cite{Anderson56} Magnetite is very familiar to a large cross-section of the condensed matter community as the archetype of strong correlation physics predating the cuprate superconductors by many decades. 

The Coulombic charge ice model, it turns out, does not adequately account for the charge ordering transition of magnetite and, indeed, the low temperature physics of this material is not fully understood to this day - around $70$ years after the original work (see, e.g., Refs.~[\onlinecite{Walz2002,Senn2011}]). The principal difficulty in making theoretical headway on magnetite is that many different degrees of freedom - spin, orbital, structural, insulating and itinerant - are coupled so that they cannot be considered in isolation. In this paper, we revisit the Coulombic charge ice model of Verwey and Anderson in the light of recent progress on the spin ices and Coulomb phases in general. We have conducted a detailed study of the thermodynamic properties of Coulombic charge ice with the aim of making explicit phenomenological predictions from the model which may lead to  experimental evidence of the frustration of the Coulomb interaction in magnetite or related materials despite the competing effects of other degrees of freedom at temperatures above the charge ordering transition.

The charge ice model may be obtained from a model of spin polarized fermions on a pyrochlore lattice at half-filling with nearest neighbor repulsion.\cite{Fulde02,PhysRevLett.97.170407} This nearest neighbor model, in the classical limit, exhibits the same ground states as nearest neighbor spin ice - these states have a one-to-one correspondence. Coulombic charge ice is then obtained from the nearest neighbor model by endowing the fermions with electric charge in the presence of a neutralising background charge. 
Including long-range Coulomb interactions to the model of mobile fermions on a pyrochlore lattice is analogous to adding dipolar interactions to the nearest neighbor spin ice model. The latter was a crucial step towards a full understanding of the spin ice materials. In one sense, this step makes the physics less clean since the extensive degeneracy of the nearest neighbor interacting model is broken. Remarkably the Coulomb phase in the ice model not only survives the addition of a long range dipolar interaction - the degeneracy is only weakly broken owing to the screening of the long range interaction - but also the dipolar interaction between defects about the ice states is fractionalized leading to an effective $1/r$ Coulomb interaction between these defects. \cite{Castelnovo2011} We will show that the connection between the Coulombic charge ice model and the dipolar spin ice model is, in fact, more than an analogy and that insights from one model allow us to draw conclusions about the other. 

We organize the paper as follows. We continue the introduction (section~\ref{sec:Fe3O4}) with a short review of dipolar spin ice physics highlighting the points relevant to our study. This is followed in ~\ref{sec:Fe3O4} by a short description of some aspects of magnetite phenomenology. For those familiar with both we recommend beginning with Section~\ref{sec:CCI} in which we begin (section~\ref{sec:energetics}) by showing how the dipolar spin ice model and the Coulombic charge ice models may be treated on an equal footing in a sense to be made precise later. We also discuss the ground states of the two models. Then, in Section~\ref{sec:MC}, we explore the thermodynamics of charge ice via classical Monte Carlo simulations. A number of the features that we observe in these simulations can be understood within the framework of large $N$ mean field theory (Section~\ref{sec:spectrum}) including the screening of the long range interactions and the ordering wave vector. A more detailed exploration of the connections between the dipolar spin ice model and the charge ice model may be found in three appendices. In Section~\ref{sec:cc}, we study the finite temperature charge-charge correlations and then the excitations above the ice states with the nontrivial result that the elementary defects interact via an effective $1/r$ potential (Section~\ref{sec:monopole}).  We examine the effects of a static uniform electric field on the Coulomb phase in Section~\ref{sec:EField}. In Section~\ref{sec:pseudogap}, we pinpoint three different transport regimes including a poor insulator regime existing over a broad temperature range encompassing the Coulomb phase. Finally, we summarize our results and consider the outlook for further studies of frustrated charge models.

\subsection{Dipolar Spin Ice}
\label{sec:DSI}

In this section, we give a brief introduction to the spin ices concentrating on topics that are pertinent to our discussion of charge ice. For further information see, for example, Ref.~[\onlinecite{FrustratedMagnetismBook}]. The tell-tale signatures of unconventional magnetism in the materials Dy$_{2}$Ti$_{2}$O$_{7}$ and  Ho$_{2}$Ti$_{2}$O$_{7}$ were careful measurements of the residual magnetic entropy at low temperatures obtained from specific heat data.\cite{Ramirez1999} The residual entropy was found to be, within errors, equal to the Pauling entropy of water ice. \cite{Pauling1935} The microscopic explanation for this phenomenon is as follows. The spins, which are located on the vertices of the pyrochlore lattice of corner-sharing tetrahedra, have an Ising anisotropy that forces each spin $\mathbf{S}_i$ to point along an axis connecting the centers of the two tetrahedra to which it is connected. An extensive ground state entropy in this Ising system can be obtained via nearest neighbor interactions of the form
\begin{equation}
H_{\text{nn}} = - J \sum_{\langle i,j \rangle}S_i S_j 
\label{eqn:nnsi}
\end{equation}
where the spins $S_{i}$ are written in the local Ising frame and therefore assume values $\pm1$. This interaction is frustrated and it's easy to see that the ground states are those for which $\sum_{\rm tetra} S_i =0$ - the sum of spins over each tetrahedron vanishes - this is the so-called ice rule and states fulfilling this condition are the spin ice states. This local constraint on the tetrahedra translates to an extensive degeneracy on the lattice with a Pauling entropy. The local ground state constraint can be coarse-grained by introducing fields $\mathbf{H}(\mathbf{r})$ to denote the dipolar orientations. The local constraint is then $\nabla\cdot\mathbf{H}=0$, so the field has only a circulation and no sources. This constraint and the extensive entropy can be shown to imply dipolar spin correlations at zero temperature which persist on length scales smaller than the correlation length at finite temperature.\cite{Henley2005} These correlations manifest themselves as pinch points in the structure factor centered on reciprocal lattice vectors.\cite{Henley2005}

\begin{figure}[h!]
\includegraphics[width=9cm]{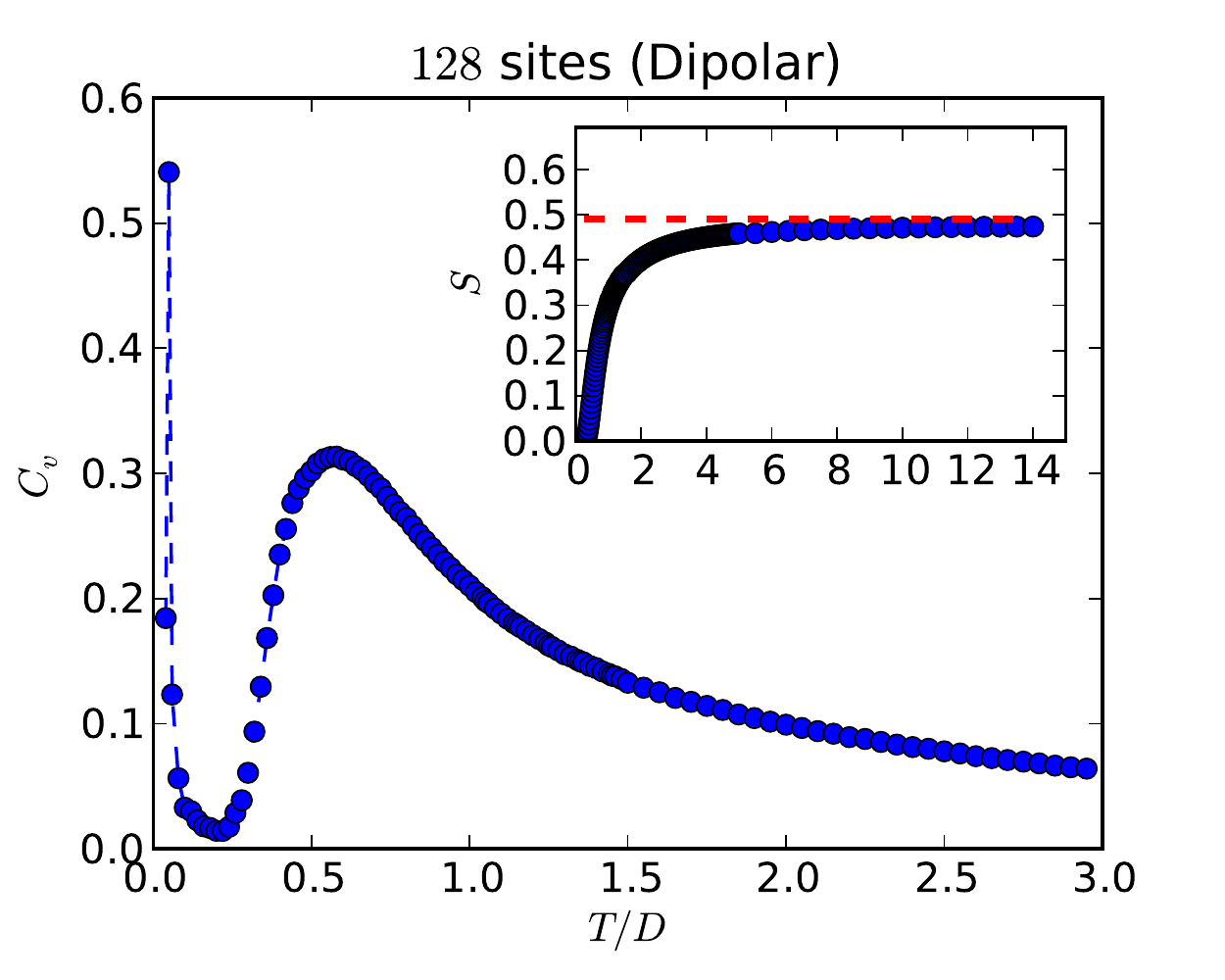}
\caption{Monte Carlo results on a 128-site cluster for the specific heat of the  spin ice model which only takes into account the dipolar term. Inset: Entropy release above $T_C$. The dashed line shows the Pauling estimate $\approx \ln2 - \frac12 \ln\frac32$.\label{fig:cv_d}}
\end{figure}

In the spin ice materials, the situation is, at least at first sight, complicated by the fact that the dominant interactions between the moments are long-range dipolar interactions. Indeed, the model appropriate to the spin ice materials is the so-called dipolar spin ice model with Hamiltonian
\begin{eqnarray}
H_{\text{dipolar}}&=& - J \sum_{\langle i,j \rangle}\mathbf{S}_i \cdot \mathbf{S}_j  \\
&+&D \sum_{ij}\left[\frac{\mathbf{S}_i \cdot \mathbf{S}_j}{|\mathbf{r}_{ij}|^3}-\frac{3(\mathbf{S}_i \cdot \mathbf{r}_{i})(\mathbf{S}_j \cdot \mathbf{r}_{j})}{|\mathbf{r}_{ij}|^5}\right]\nonumber
\label{eq:dip}
\end{eqnarray}
Experimentally, one observes a crossover into a spin ice regime as the temperature is lowered. Within this regime, all but the two-in/two-out spin ice states are frozen out. The dipolar spin ice model accounts quantitatively for the much of the phenomenology of the spin ices. Despite the, albeit weak, breaking of the degeneracy of the spin ice states by the long-range dipolar interaction, the microscopic picture presented above for the nature of the spin ice regime remains intact.\cite{Melko2001, Melko2004, Isakov2005} The one notable discrepancy between the theory and the experiment is that the model predicts a first order transition to a long-range ordered phase within the spin ice regime. \cite{Melko2001,Melko2004} This transition shows up in Fig.~\ref{fig:cv_d} as a sharp peak (see Sec.~\ref{sec:MC} for detail on the numerical simulation).The transition has not been observed in the spin ice materials presumably because the spin ices fall out of equilibrium at sufficiently low temperatures (though see Ref.~[\onlinecite{Pomaranski}]). 

Let us consider the effect of carrying out a single spin flip in the spin ice regime. This produces a pair of tetrahedra that violate the spin ice rule. By carrying out successive spin flips, the defected tetrahedra can be spatially separated. In the language of coarse-grained variables, the divergence-free condition on $\mathbf{H}$ is broken locally. In the electrostatic analogy mentioned above, we would understand $\nabla\cdot\mathbf{H}\neq 0$ to reflect the existence of a charge. One can show that the analogy extends to an effective interaction between the defected tetrahedra. In the nearest neighbor model, Eq.~(\ref{eqn:nnsi}), the defects interact via a Coulomb interaction of entropic origin.\cite{Henley2005} Shortly after the entropic interaction was noted, it was observed that there is an energetic effective interaction. Remarkably, the dipolar interactions themselves fractionalize into Coulomb interactions between effective charges centered on the defected tetrahedra. \cite{Castelnovo2008} One of the implications of the effective Coulomb interactions is that, over some temperature range where the concentration of defects is small and the charges of the defects have the same magnitude, the thermodynamics of the spin ice materials should be identical to that of a dilute electrolyte to a good approximation. These insights into the nature of the spin ice state and the excitations about it have led to a number of experimental works exploring aspects of this physics through neutron scattering \cite{Kadowaki2009,Morris2009,Fennell2009}, $\mu$SR \cite{Bramwell2009,Dunsiger2011} and bulk measurements.\cite{Morris2009,Castelnovo2011,Giblin2011}

\subsection{A Very Brief Introduction to Magnetite}
\label{sec:Fe3O4}

Magnetite is an inverse spinel at room temperature with two inequivalent crystal sites for the iron ions and an fcc structure for the oxide ions. \cite{Anderson56} Merely counting oxidation states suggests that both Fe$^{2+}$ and Fe$^{3+}$ are present. X-ray refinement and evidence from other measurements led to the conclusion that half of the Fe$^{3+}$ ions occupy the tetrahedral A sites while equal numbers of Fe$^{2+}$ and Fe$^{3+}$ occupy the pyrochlore $B$ sites. The material is a ferrimagnet below $848$ K and exhibits a further transition at $125$ K which has been named after Verwey. \cite{Verwey1947} The low temperature transition and the nature of the phases that it connects have been the subject of debate for many decades. \cite{Walz2002} The Verwey transition is known to be first order and is coincident with a jump in the resistivity upon cooling and a structural change into a low temperature monoclinic structure. The low temperature phase appears to be charge ordered and many studies have attempted to pin down the ordered structure. One idea that dates back to Verwey and which was followed up by Anderson is that the charges are predisposed to undergo short range ordering into a state with predominantly two Fe$^{3+}$ and two Fe$^{2+}$ on each tetrahedron in one-to-one correspondence with the magnetic spin ice states. The Verwey transition was attributed to long-range order emerging from the ice-like order. \cite{Anderson56} This picture now appears to be greatly oversimplified not least because recent refinements of the low temperature charge order indicate the the ice rule is not satisfied. \cite{Senn2011} For further details, we refer to the review Ref.~[\onlinecite{Walz2002}] and recent X-ray refinements of the charge order structure Ref.~[\onlinecite{Senn2011}] and references therein. However, our interest in this model is not merely historical: we suggest that the frustration of the Coulomb interaction may lead to clear experimental signatures in magnetite or related materials.

\section{Coulombic Charge Ice}
\label{sec:CCI}
In order to study the effects of long range interaction on the physics of charge ice, we consider the model Hamiltonian
\begin{equation}
H_{\text{Coulomb}}=V\sum_{ij}\frac{q_iq_j}{|\mathbf{r}_{ij}|}.
\label{eq:cou}
\end{equation}
Here the vector $\mathbf{r}_{ij}$ connects the charges $q_i=\pm 1$ which are sitting on the sites of the pyrochlore lattice.  The analogy between charge model and the dipolar spin ice  arises due to the fact that the low-lying energy states for both models constitute a manifold in which each state obeys local `ice rules'. Specifically, in the charge case, the ice rules translate into have two $q=+1$ and two $q=-1$ charges on each tetrahedron. In the dipolar case, there are always two `in' and two `out' spins for a given tetrahedron. Through comparison of the two model Hamiltonians we can investigate the relationship between the degrees of freedom in each case.

\subsection{Multipole Expansion for the Energetics of Spin Ice and Charge Ice}
\label{sec:energetics}

One role of the long range interactions in dipolar spin ice and Coulombic charge ice is to lift the degeneracy of the ice rule states. Since the interactions are very different in the two models we would not, as first sight, expect the degeneracy breaking to have any commonalities. For example, the ground states of the models are different. The central result of this section is that from our two distinct long-range interacting frustrated models in three dimensions, we can write down a third model from which our two ice models are descended. From this ``ancestor" model, we can identify qualitative features relating the spectra of charge ice and dipolar spin ice.

\begin{figure}[h!]
\includegraphics[width=8cm]{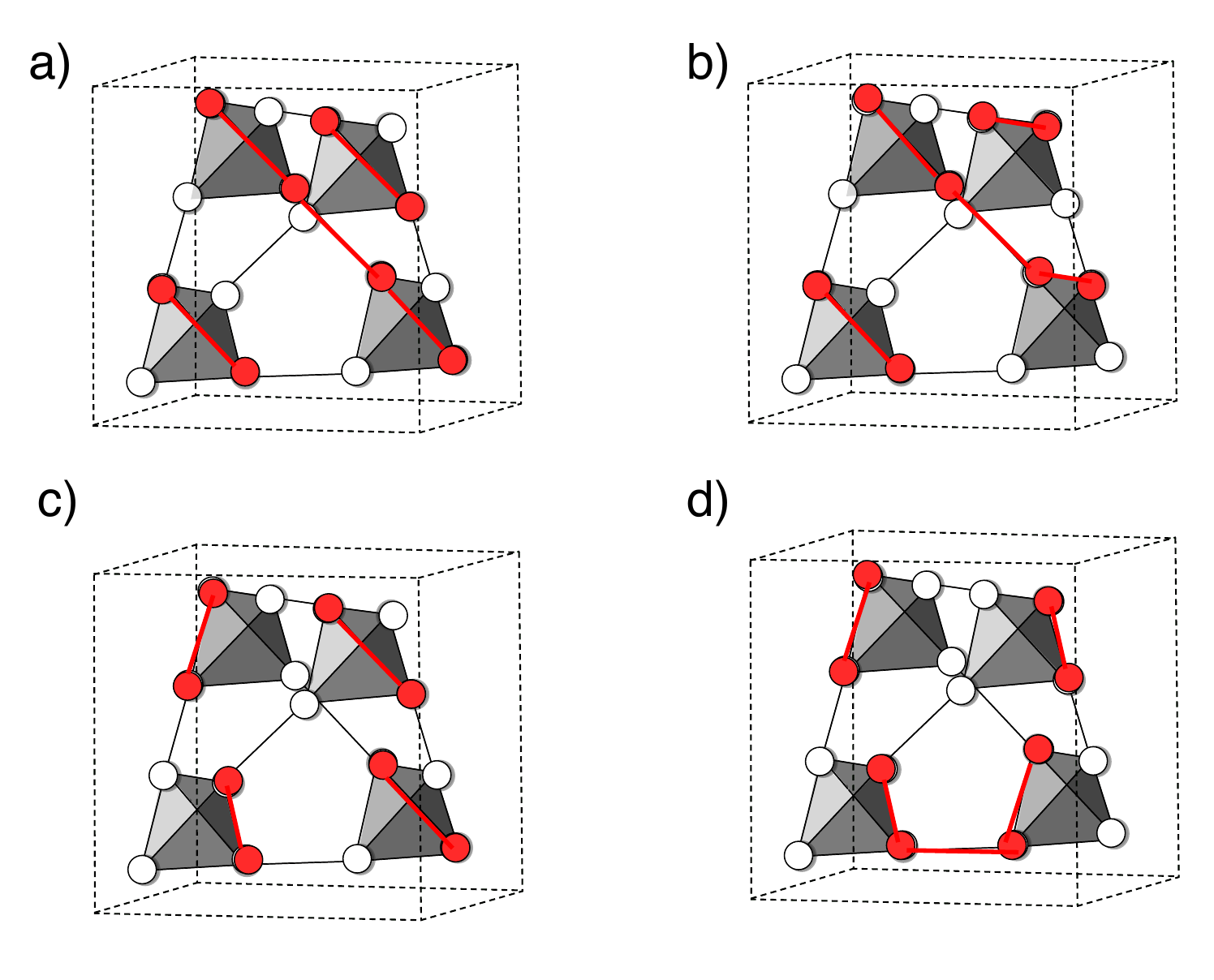}
\caption{The four inequivalent ice states which are compatible with the cubic 16-site unit cell. Panel (a) illustrates the ground state of the long-range interacting charge ice model and panel (d) represents a unit cell of the dipolar spin ice ground state.
\label{FIG:4STATES}}
\end{figure}

\begin{figure}[h!]
\includegraphics[width=8cm]{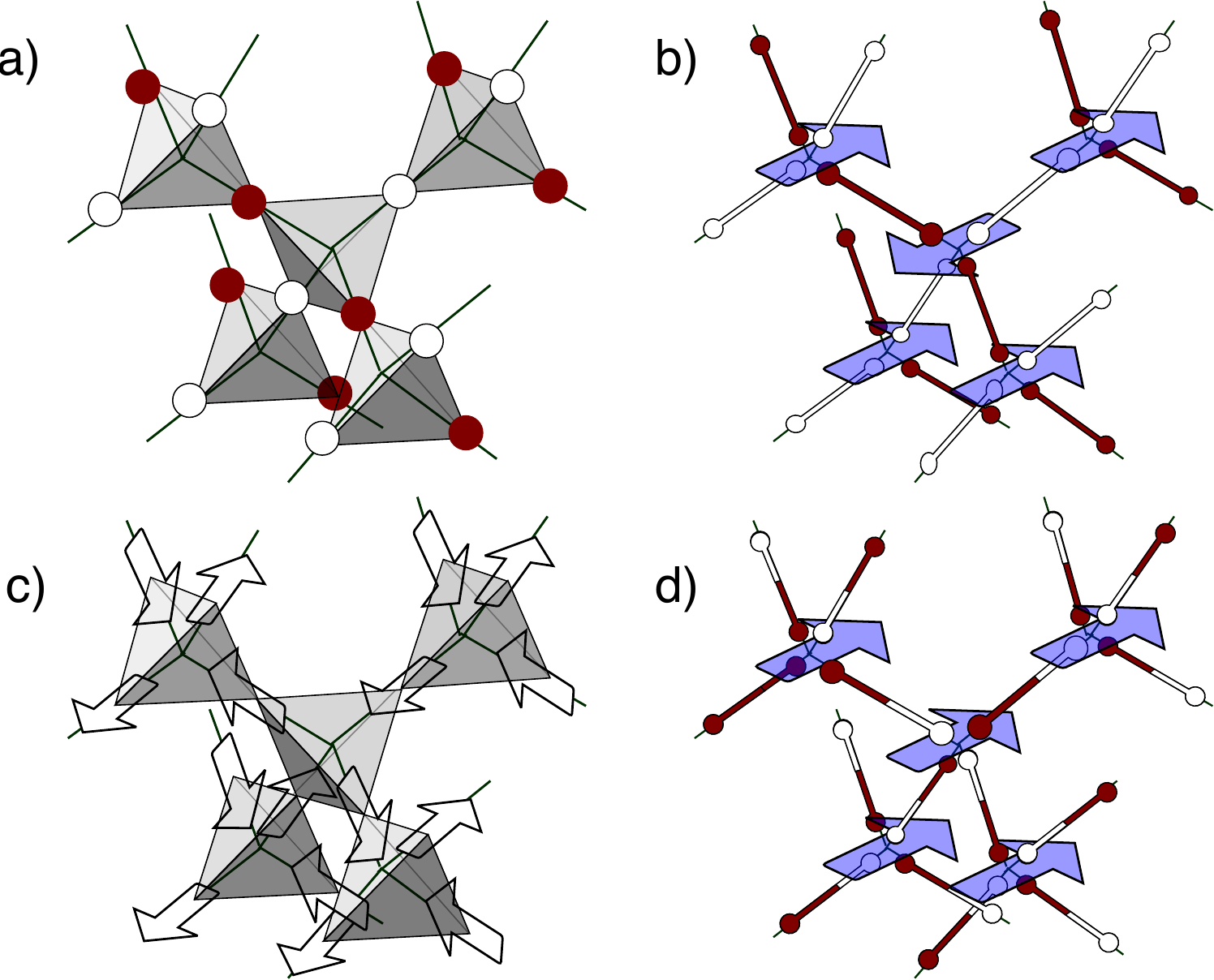}
\caption{\label{fig:dumbbell} Mapping of the dipolar and charge model to dumbbells. For the charge model, consider panels (a) and (b) which show five tetrahedra within a pyrochlore structure with the vertical direction $[001]$. We break each of the charges $q$ in panel (a) into a pair of charges $q/2$ separated by distance $l$ with the line joining the smaller charges connecting the centers of the two neighboring tetrahedra. The pyrochlore charge configuration on the left maps to the charge configuration on the right. Now, each set of four $q/2$ charges within each tetrahedron has a set of multipole moments. The ice configurations have zero charge on each tetrahedron but they do have a net dipole moment. The dipole moment is illustrated as an arrow centered on each diamond site. For the dipolar model, we refer to panels (c) and (d). Here we break the dipole moments on the pyrochlore sites into a pair of equal and opposite charges. Again, the line joining the charges connects the centers of the two neighboring tetrahedra. Once again, grouping the charges together on each tetrahedron gives a model of interacting multipole moments.}
\end{figure}

Before we do this, we describe the ground states of the two models. The lowest energy state of the charge ice model is a state with charges of identical sign along $[110]$ chains of the pyrochlore lattice and alternating charges on perpendicular $[1\bar{1}0]$ chains. This charge configuration is illustrated in panel (a) of Fig.~\ref{FIG:4STATES} and is one of the four ice states of distinct energy which are commensurate with the smallest cubic unit cell and which are illustrated in the same figure. The dipolar spin ice model ground state, in contrast, has a ground state with ordering wavevector $\mathbf{q}=(0,0,2\pi)$ which can be mapped into the state of panel (d) of Fig.~\ref{FIG:4STATES}. \cite{Melko2001,Melko2004} The mapping is described in Fig.~\ref{fig:dumbbell}. This is the highest energy ice state of the charge model compatible with the $16$ site cell. 

The effect of frustration of the Coulomb interactions was pointed out by Anderson who computed the Coulomb energies for the ground state charge configuration and a second long-ranged order state which is energetically disfavored by having neighboring $[110]$ chains with alternating charges in phase. From the energy difference for these two states, he estimated that the energy scale for long range order is about $5\%$ of the characteristic energy of the Coulomb interaction for charges on a pyrochlore lattice. This suppression of the scale for long-range order well below that of the scale of interactions is vividly displayed in the spin ice materials.

We map from the charge model to the spin model by taking a charge $+1 (-1)$ (filled (open) circle) to an inwards (outwards) pointing dipole on those tetrahedra centred on fcc lattice points. Thus panel (c) in Fig.~\ref{fig:dumbbell} is obtained by a mapping from panel (a). We note that while all charge model states can be mapped into dipolar model states, the reverse is not true. This difference arises from the fact that the charge model is constrained to half filling while the dipolar model has no such constraint. For example, there is no analogue in the charge model of a pair of spin states differing by a single spin flip without adding or removing a charge. 

As advertised above, the charge ice and spin ice models can be treated on a similar footing. This is achieved by splitting the degrees of freedom of the two models into charges associated with the lattice formed by the centers of tetrahedra. The idea behind this {\it dumbbell picture} is described in Fig.~\ref{fig:dumbbell}. \cite{Castelnovo2008}  The splitting of the system into clusters of charges and considering a multipole expansion has been carried out for artificial spin ice arrays where the elementary dipoles are mesoscopic magnetised ``islands". \cite{Moller} When we make the replacement of all microscopic degrees of freedom by dumbbell charges, we notice that the ice states (b) and (c) in Fig.~\ref{fig:dumbbell} are not identical. The charges on the central tetrahedron have opposite signs. 

Now we consider the four charge cluster around each diamond lattice site (the sites formed by the centers of the tetrahedra). The cluster can be characterised by a set of multipoles. In the ice manifold, the net charge on each cluster is zero. The dipole moment of an ice state may have components along any of the cubic cell axes of Fig.~\ref{FIG:4STATES}. In particular, if the cubic cell edge length is $L$ then the distance between neighbouring tetrahedra is $a\equiv  \sqrt{3}L/4$. Introduce parameter $\epsilon \equiv 1- \frac{l}{a}$ where $l$ is the dumbbell length which varies between $0$ in the diamond lattice limit and $1$ in the pyrochlore limit. The charges $q$ on pyrochlore sites are split in $q/2$ charges on the dumbbells. The dipole moments are then $\frac{q\epsilon L}{4}\left( 0,0,1 \right)$ with others related by cubic symmetry. The quadrupole moments are given by
\[  Q_{\alpha\beta} = \frac{1}{2} \sum_{i} q_{i}\left( 3r_{i\alpha}r_{i\beta} - r_{i}^{2}\delta_{\alpha\beta}  \right)  \]
with six Cartesian components of which only one is non vanishing for the ice states. For example $Q_{xy} = 3q\left( \frac{\epsilon L}{8} \right)^{2}$ for one of the ice states. Since the pyrochlore may be split into bipartite A and B tetrahedra, the charge ice states and dipolar spin ice states are related by having identical multipole moments on the A tetrahedra and flipped moments on the B tetrahedra or vice versa. The leading interactions are those between dipole moments because interactions between higher multipoles fall off as a higher power of their separation. These observations lead us to consider a model of dipoles on a diamond lattice where the dipoles can point in any of the $[001]$ crystallographic directions. Owing to the staggering between dipole moments between the charge ice and dipolar spin ice models, we generally have to impose an extra constraint on the dipolar orientations. When the constraint is imposed, it turns out that the dipolar interactions between diamond lattice sites are sufficient to obtain the ground states of the charge ice and dipolar spin ice models.  It is interesting, however, to relax this constraint and to find the ground state configuration of the model when all the dipoles can assume any of their six configurations. We find that the ground state is the analogue of the $12$ fold degenerate $\mathbf{q}=0$ ground state that is also the ground state of the dipolar spin ice model \cite{Melko2001,Melko2004} so, at least for the ground state, the diamond lattice model imposes the ice constraint energetically. In Section~\ref{sec:spectrum} and Appendix~\ref{appx:2}, we further use this diamond lattice ``ancestor" model of dipoles to see explicitly in reciprocal space that properties of the charge and dipolar models are descended from it via a projection. 

Another observation we can make from the dumbbell decomposition and the staggering between the two models is that there should be an anticorrelation between the ice spectra of the two models. In other words, the highest energy ice states of one model should be among the lowest of the other and vice versa. In Appendix~\ref{appx:1}, we show that the spectra of the ice states are, in fact, exactly inverted for a $16$ periodic cell and, in larger clusters, while the inversion ceases to be exact, we find evidence for the anticorrelation of energies.

\subsection{Monte Carlo results}
\label{sec:MC}

We perform Monte Carlo simulations of the charge ice model and compare the results to the ones found for spin ice. The simulations are carried out on $L^3$ pyrochlore clusters of cubic symmetry with periodic boundary conditions ($L$ measures the linear size of the cubic cluster with respect to a 16-site unit cell). Long range interactions are taken into account using an Ewald summation. The Ewald summation method is one of the most extensively developed numerical techniques for the computation of long-range interactions. It recasts a conditionally convergent series (such as that of the Coulomb energy of a charge neutral system), into two rapidly converging series in real and reciprocal-space respectively.\cite{Board1996}  Hence, using this method it is possible to take into account the full Coulomb interaction, thereby importantly preserving subtle manifestations of long-range interactions, such as screening effects discussed in the Introduction.

We use swapping updates (which simply exchange two opposite charges) to calculate the specific heat.  As expected, the acceptance rate of these updates decreases dramatically at low temperatures. We therefore additionally employ a Metropolis algorithm with non-local updates, so-called `worm' updates, to calculate energies of the model system. These updates allow us to effectively bypass the energy barriers that separate nearly degenerate states and thus to sample the low energy states more efficiently. \cite{Evertz2003,Melko2001,Melko2004} In a `worm' update, two initially neighboring sites of opposite charge are exchanged which creates a pair of defects. These defects are considered to constitute the two ends of the worm (or string). They move around the lattice system until they `annihilate' each other; that is, they meet on the same tetrahedron. The update therefore does not create additional defects of the ice rule. 

\begin{figure}[h!]
\includegraphics[width=9cm]{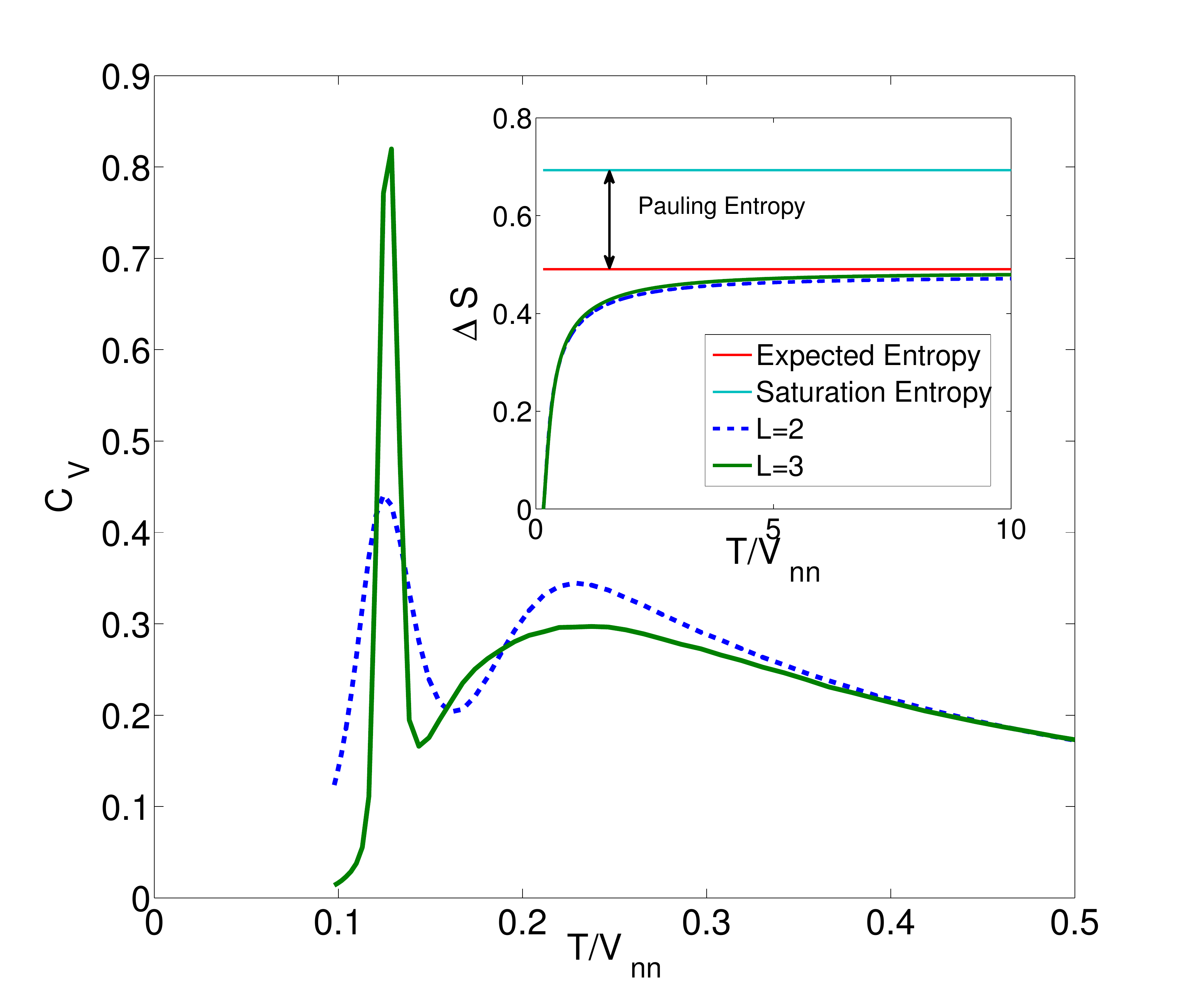}
\caption{Monte Carlo results for the Coulombic charge model. The specific heat of the 128-site ($L=2$) and 432-site ($L=3$) clusters is shown in the main panel. Inset: Entropy release above $T_C$ for the Coulomb model. The lower horizontal line shows the Pauling estimate $\approx \ln2 - \frac12 \ln\frac32$ and the upper line shows the saturation entropy of $\ln 2$.\label{fig:cv_s}}
\end{figure}

\begin{figure}[h!]
\includegraphics[width=9cm]{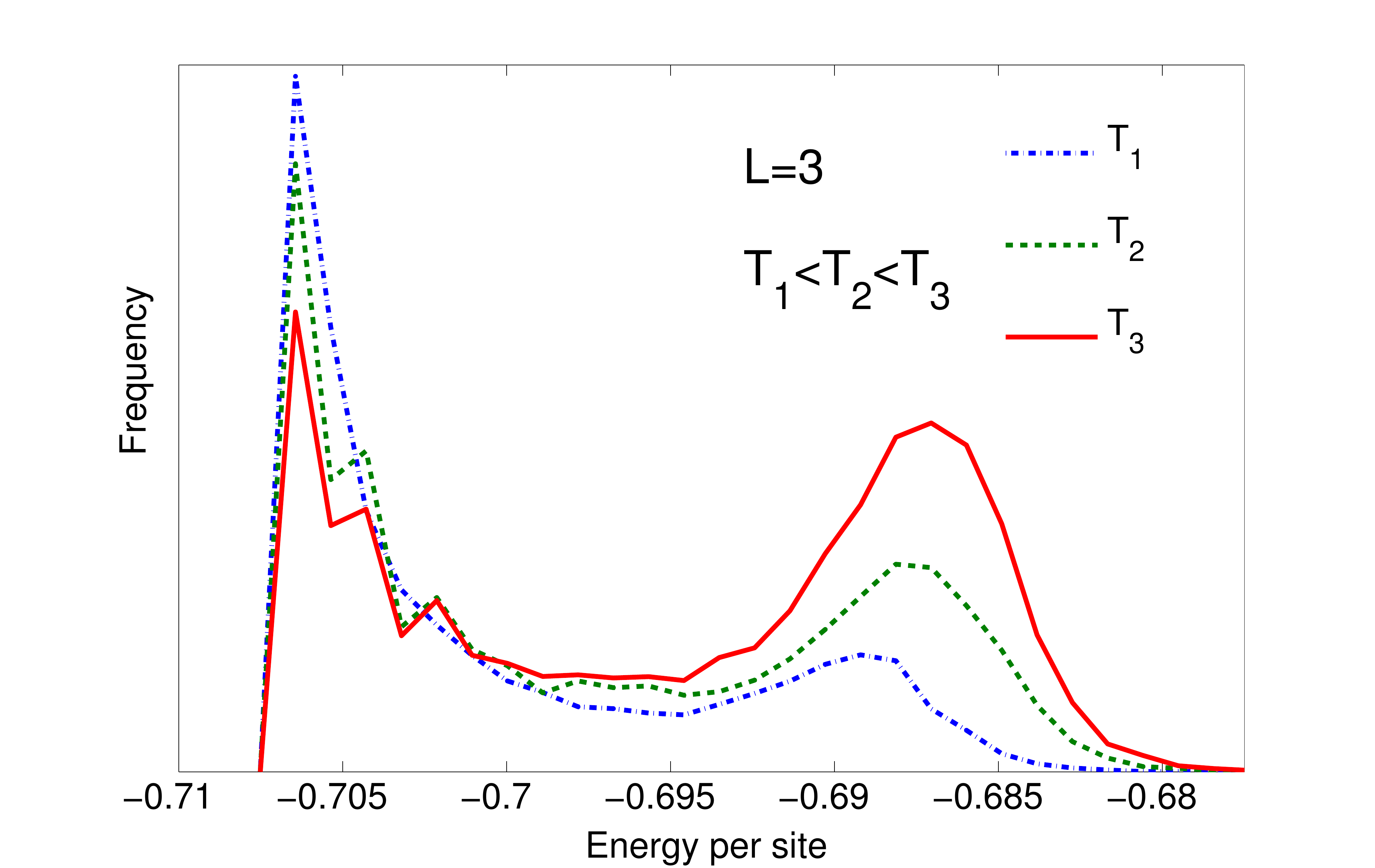}
\caption{Figure showing the histogram of Monte Carlo sampled energies for three temperatures in the vicinity of the finite size transition temperature. The double peak indicates that the transition is first order. \label{fig:histogram}}
\end{figure}

We focus on the specific heat.  For the 16-site case it is possible to compare the specific heat values calculated from the Monte Carlo simulations with the exact analytical calculation; they are seen to be in perfect agreement.  Fig.~\ref{fig:cv_s} shows Monte Carlo data for larger system sizes: $L=2$ ($128$ sites) and $L=3$ ($432$ sites). A broad peak in the specific heat indicates  a crossover into to the ice-manifold (i.e., the point at which all defects disappear). At low temperatures a second, sharp peak results from a first order phase transition from the Coulomb phase into an ordered phase (the $T=0$ ground-state of the Coulomb model is a 6-fold degenerate state shown in Fig.~\ref{FIG:4STATES}a). A numerical integration of the specific heat divided by the temperature is performed, giving the entropy change from the charge ice regime into the high temperature phase. The relation used to obtain the entropy is
\begin{equation}
S(T_2)-S(T_1)= \int_{T_1}^{T_2} \frac{C(T)}{T} dT.
\end{equation}
We choose $T_1$ to be slightly above the ordering temperature and $T_2$ is chosen to be deeply in the paramagnetic (charge disordered) regime. This entropy can be compared with Pauling's estimate for the residual entropy in water ice \cite{Pauling1935, Pauling}, due to the analogy between the degrees of freedom of the spin ice models and those of water ice. \cite{Melko2004} The calculations indicate that at temperatures above the transition into the ice manifold, the entropy release is in good agreement with the Pauling estimate, i.e.,  $\Delta S \approx  \ln2-\frac{1}{2}\ln\frac{3}{2}$. 

Fig.~\ref{fig:histogram} shows histograms of energies obtained from Monte Carlo simulations of a $432$-site ($L=3$) cluster. Data is shown for three temperatures taken close to the low temperature heat capacity peak which signals the onset of charge order. The double peak in the histograms indicates that the transition is first order.

\section{Mean Field Theory}

\subsection{Spectrum of Interactions}
\label{sec:spectrum}

A very useful way of studying spin ice physics is to look at the spectrum of the interactions in reciprocal space. \cite{denHertog2001, Isakov2005} If we write the Hamiltonian coupling the spins as $H=\sum_{\mathbf{k},a,b}\mathcal{J}_{\mathbf{k}}^{ab}S_{\mathbf{k}}^{a}S_{-\mathbf{k}}^{b}$ where $a$ and $b$ are the sublattice labels, then one can show that, of the four eigenvalues of $\mathcal{J}_{\mathbf{k}}^{ab}$, the lowest two are degenerate over the Brillouin zone and almost flat compared to the overall splitting of the spectrum. The upper two bands are strongly dispersive. One can make various connections between this spectrum and the properties of dipolar spin ice known from Monte Carlo simulation. 

The most profound connection lies precisely in the flatness of the two lowest bands. \cite{denHertog2000,Isakov2005} If the long-ranged dipolar coupling is truncated beyond nearest neighbor sites on the pyrochlore lattice, the two lowest bands are perfectly flat and reflect the degeneracy of the two-in/two-out spin configurations. With the full long-ranged dipolar coupling, the degeneracy of the lowest bands is lifted but sufficiently weakly that for temperatures greater than the bandwidth of these bands, the ice states are effectively degenerate and defects out of the ice manifold are exponentially small in the temperature. Surprisingly, the breaking of the ice band degeneracy takes place on a smaller energy scale than the scale of the second neighbor coupling. \cite{denHertog2000,Isakov2005} In addition to the robustness of the ice states, one can also determine the ordering wavevector of the transition to long range order that happens within the spin ice state - this wavevector is simply the location in reciprocal space of the lowest eigenvalue within the lowest pair of bands. 

\begin{figure}[h!]
\includegraphics[width=9cm]{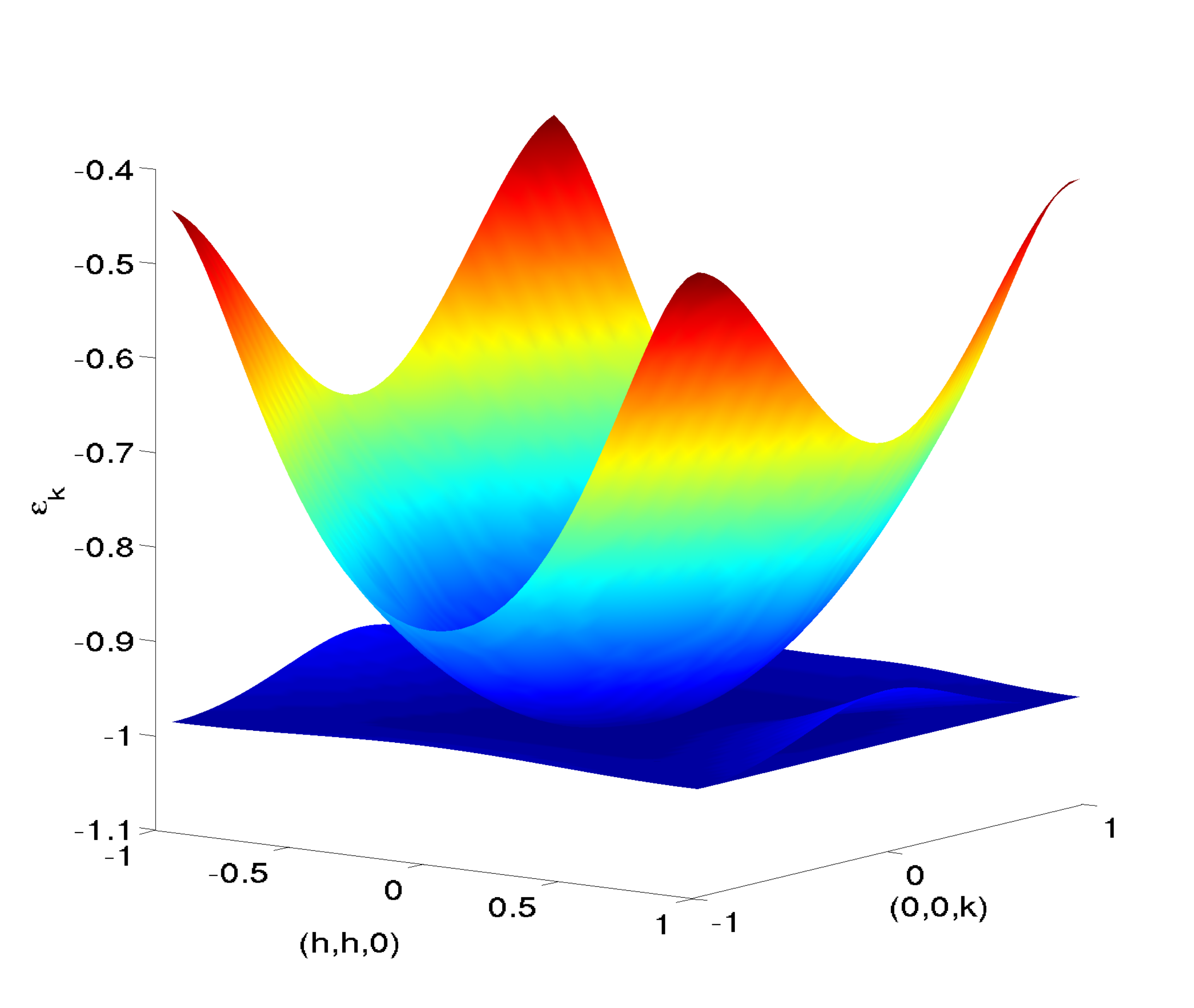}
\caption{\label{fig:flat_band} Lowest three bands in the spectrum of $V_{\mathbf{k}}^{ab}$ for the charge ice model. The eigenvalues of $\tilde{V}_{\mathbf{k}}^{ab}$ are plotted in the $hhl$ plane in reciprocal space. The fourth band has a divergence and is omitted.}
\end{figure}

We can repeat this analysis for the Coulomb interaction in charge ice. The Hamiltonian can be written as $H=\sum_{\mathbf{k},a,b} \tilde{V}_{\mathbf{k}}^{ab}Q_{\mathbf{k}}^{a}Q_{-\mathbf{k}}^{b}$. The spectrum of the Coulomb interaction $V_{\mathbf{k}}^{ab}$ once again consists of two almost flat bands compared to the total bandwidth of the third band (Fig.~\ref{fig:flat_band}) and the flatness of the lowest bands is descended from the geometrical frustration of the nearest neighbor interactions on the pyrochlore lattice. Unlike the dipolar model spectrum, the Coulomb interaction contains a $q^{-2}$ divergence which appears only in the highest band and which arises because the interaction does not, on its own, enforce charge conservation. 

The position of the minimum eigenvalue in reciprocal space gives the same ordering wavevector of charge ice that is observed in Monte Carlo simulations, namely $\mathbf{k}=0$ as opposed to $\mathbf{k}=001$ in spin ice. The principal differences in the spectrum compared to the case of dipolar spin ice are (i) one of the bands of the charge ice model has a divergence at $\mathbf{q}=0$  that the spectrum of dipolar spin ice has a gap between the lowest two (ice) bands and the third band. Charge ice, in contrast, is gapless at the $\Gamma$ point. We can understand this on the basis of the number of eigenvectors at the wavevectors of minimal eigenvalue which provide information about the number and nature of the ordered states - a fact which forms the basis of the Luttinger-Tisza method \cite{LuttingerTisza} for finding ground states of spin systems. One reason for the absence of the gap in charge ice is that the ground state at $\mathbf{k}=0$ is six-fold degenerate at the $\Gamma$ point. With only two degenerate bands at $\mathbf{k}=0$, the degeneracy can be at most four-fold including time reversal and a third touching band is required to obtain all six states in the ground state. This argument is also applicable to nearest neighbor spin ice which also has a touching point in its spectrum at the $\Gamma$ point. For dipolar spin ice, the ground states are at $\mathbf{k}=001$. The counting argument for the $12$-fold degenerate states in this case is that there are three $\mathbf{k}=001$ positions related by symmetry which is multiplied by two owing to the degeneracy of the ice bands. Time reversal brings the number of states to $12$ as required.

 In the presence of long range dipolar interactions, the eigenvalues within the lowest two bands become dispersive though the eigenstates change little from the nearest neighbor case. We may examine the corrections to the long distance dipolar interaction by studying the dispersion relations in the vicinity of the $\mathbf{k}=0$ point. The dispersion for both the charge ice model and dipolar spin ice is $\vert\mathbf{k}\vert^{2}$ close to $\mathbf{k}=0$ which corresponds to an effective $r^{-5}$ correction to the long distance dipolar interaction. 

The spectrum of the interactions in reciprocal space can be used to shed more light on the dumbbell model introduced in Section~\ref{sec:energetics}.
In Appendix~\ref{appx:2}, we use the ``ancestor" dipolar diamond lattice model to recover the spectra of the charge and dipolar models by projection onto their staggered dipolar states. In the same Appendix, we also demonstrate the consistency of the dumbbell approach by showing that the spectrum of interactions varies smoothly between the pyrochlore and diamond lattice limits as the length of the dumbbell changes. The diamond lattice model contains the same qualitative physics as the original ice models and, in principle, may be used as an approximation to them. 

\subsection{Structure Factor}
\label{sec:cc}

One of the most striking insights into the nature of spin ice is that the local divergence free constraint in tandem with the large degeneracy lead to so-called pinch points in the spin-spin correlation function. At zero temperature in nearest neighbor spin ice, these pinch points are sharp features of the correlation function and have the property that the limit as the centre of the pinch point is approached in reciprocal space, is ill-defined. In particular, in the neighborhood of the pinch point, the correlation function varies as $\delta_{\mu\nu}-\hat{q}_{\mu}\hat{q}_{\nu}$. At finite temperature, these pinch points are somewhat smeared out - an indication that there is a finite correlation length - which is related to the density of quasiparticles in the system. 

In charge ice, one might expect some signature of the Coulomb phase in the charge-charge correlation function. In this section, we compute this correlation function and compare to the case of dipolar spin ice. \cite{Sen2013} In order to carry out the computation we straightforwardly adapt a mean field theory that was developed for the Ising spin models. \cite{Enjalran2004} As we discussed in the previous section, it is useful to consider the Fourier transform of the Coulomb energy
\begin{equation}
H_{\text{Coulomb}}= \sum_{\mathbf{k}} Q_{\mathbf{k},a} \tilde{V}_{ab}(\mathbf{k})Q_{-\mathbf{k},b}.
\label{eq:cou}
\end{equation}
As above, $a$ and $b$ denote sublattices within the tetrahedral basis. The structure factor is calculated within a self-consistent mean field theory. The general idea is to soften the constraint on the charges so that one arrives at a Gaussian theory. The constraint is imposed instead through a Lagrange multiplier $\lambda$ which is computed self-consistently
\[  Q = \frac{1}{4N} \sum_{\mathbf{k}} {\rm Tr} \left[  \lambda \delta + \beta \tilde{V}(\mathbf{k})  \right]_{ab}^{-1}  \]
and we set the magnitude of the charge equal to one. The charge-charge correlation function is then
\begin{equation}
\left\langle Q_{\mathbf{k},a} Q_{-\mathbf{k},b}   \right\rangle =  \left[  \lambda \delta_{ab} + \beta \tilde{V}_{ab}(\mathbf{k})  \right]^{-1}
\end{equation}

\begin{figure}[h!]
\begin{center}
\includegraphics[width=8cm]{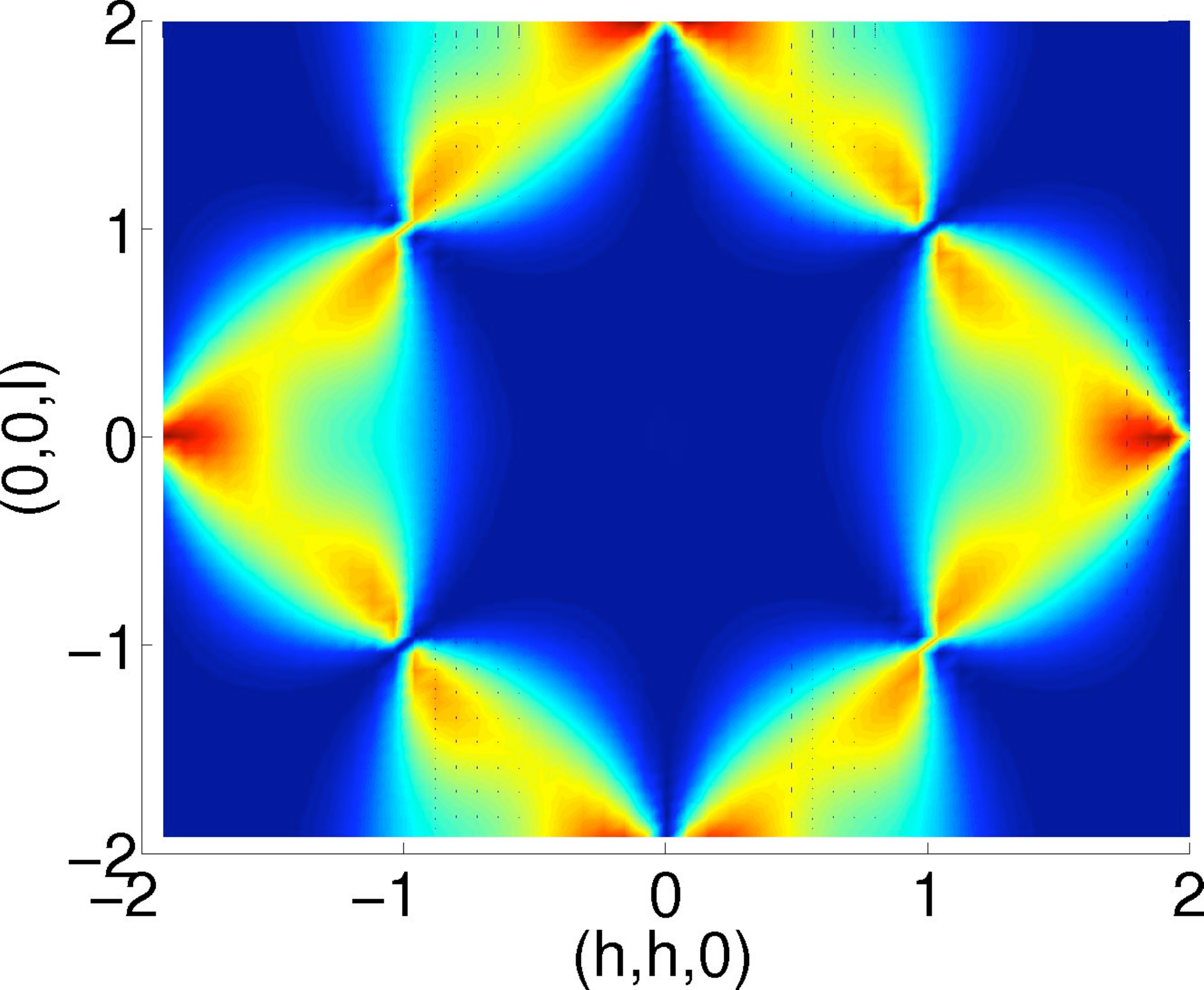}
\end{center}
\caption{Figure showing the charge-charge correlator including only weight from the bottom two bands in the spectrum of $\tilde{V}(\mathbf{k})$. \label{fig:2_T2}}
\end{figure}

\begin{figure}[h!]
\includegraphics[width=8cm]{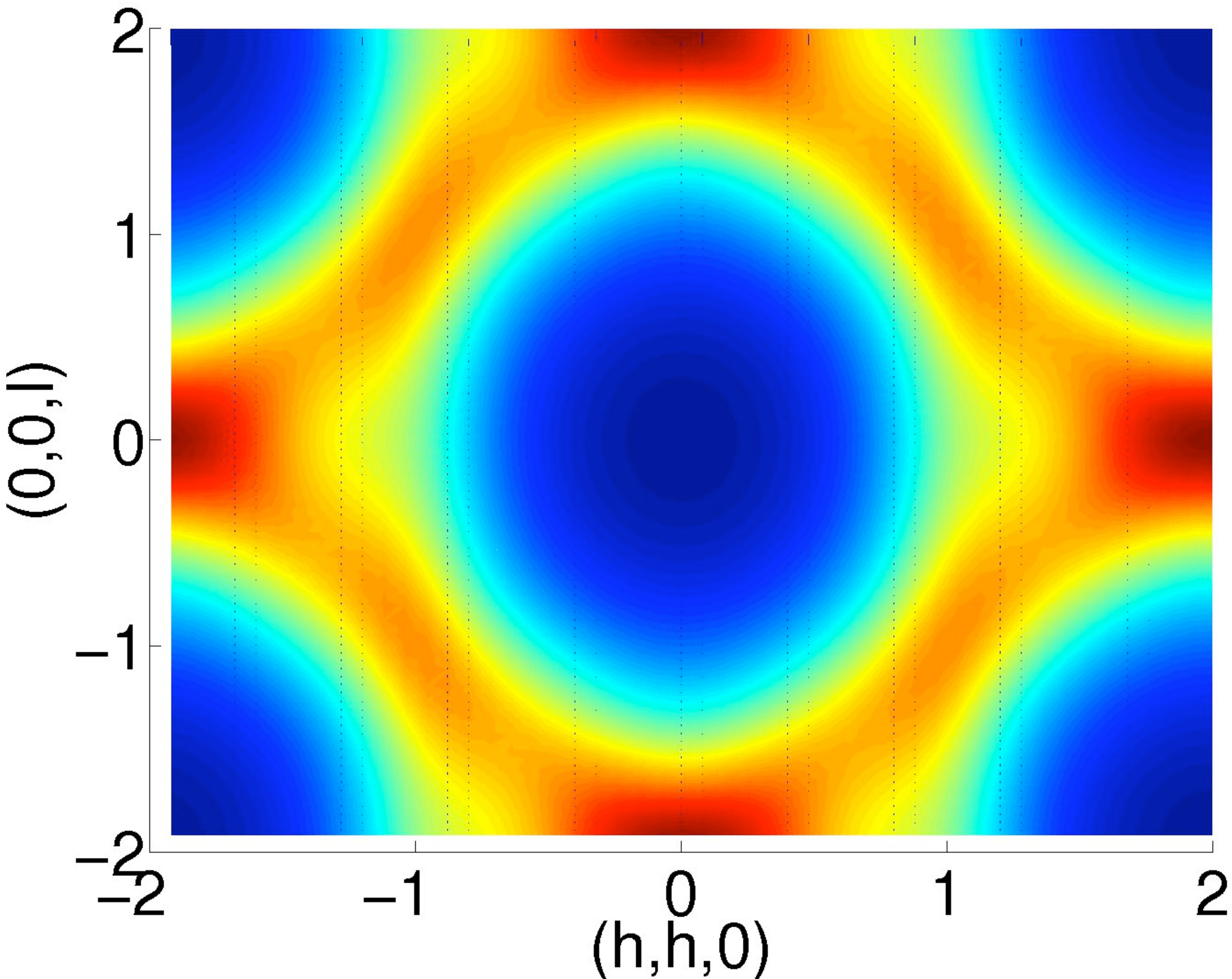}
\caption{The charge-charge correlator of charge ice at $T=0.55$ close to $T_c$. The pinch points are smeared out.}
\label{fig:QQ_finite_T}
\end{figure}

As a first step, we sum only over the bottom two bands ($\alpha=1,2$) which capture most of the weight of the ice states (in a precise sense which is laid out in Appendix~\ref{appx:3}). We obtain pinch points centred at the bcc reciprocal lattice points. The pinch point orientations are different from those in spin ice because the charge ice model maps onto a frustrated Ising model with collinear spins (Fig.~\ref{fig:2_T2}). When we include all four bands in the calculation of the correlator and set the temperature to twice the mean field transition temperature, $T=2T_{c}$, the pinch points are absent. When the temperature is lowered to $T=1.1T_{c}$, the correlations become much sharper but a clear signature of pinch points is missing (Fig.~\ref{fig:QQ_finite_T}). This result stands in contrast to nearest neighbor spin ice in which the pinch points are clearly visible at sufficiently low temperatures and dipolar spin ice in which the pinch points are present at all temperatures (at high temperatures owing to the singularity in the dipolar Hamiltonian). In terms of the spectrum of the interactions, the smearing of the pinch points is a consequence of the lowest three bands touching at $\mathbf{q}=0$ at the same time as having dispersive ice bands. In dipolar spin ice, there is a gap between the lowest two dispersive bands and the remaining bands.

The calculation of the charge correlations reveals that pinch points are not clearly resolved in the Coulomb phase even close to the charge ordering transition. If some remnant of the Coulomb phase is present in magnetite or its relatives, its experimental detection is likely to be more subtle than in the spin ices. Nevertheless, it would be a promising result to observe diffuse scattering intensity around the Brillouin zone edge with intensity modulations shown in Fig.~\ref{fig:QQ_finite_T}. 

\section{Interactions of Fractional Charges}
\label{sec:monopole}

The effective potential between a pair of point-like excitations in charge ice (in other words, in the absence of screening) has been determined by placing a pair of charges at fixed separation and measuring the energy difference after a hop of one of the excitations. This measurement was repeated a large number of times for different background ice configurations using a worm algorithm to sample the charge configuration leaving the defects fixed. The results are shown in Fig.~\ref{fig:charge_coulomb}. The periodic cell for this calculation was chosen to be $L=10$ corresponding to $16000$ sites. For our periodic cell, the correct Coulombic potential should be an Ewald potential. Indeed, the Ewald potential fits the measured potential very well. The ordinary Coulombic $1/r$ potential fits well for small charge separations compared to the system size. In dipolar spin ice, the Coulomb potential also gives an excellent fit to the measured potential (not shown) as noted previously. \cite{Castelnovo2008} The principal difference between the charge and dipolar models in this respect is that the shortest elementary defect hopping distance in charge ice is twice that in spin ice. At finite temperature, thermally excited defects are expected to screen the Coulomb potential. This introduces a length scale into the system which is reflected in the smearing of pinch points.

\begin{figure}[h!]
\subfigure[]{
\includegraphics[width=7cm]{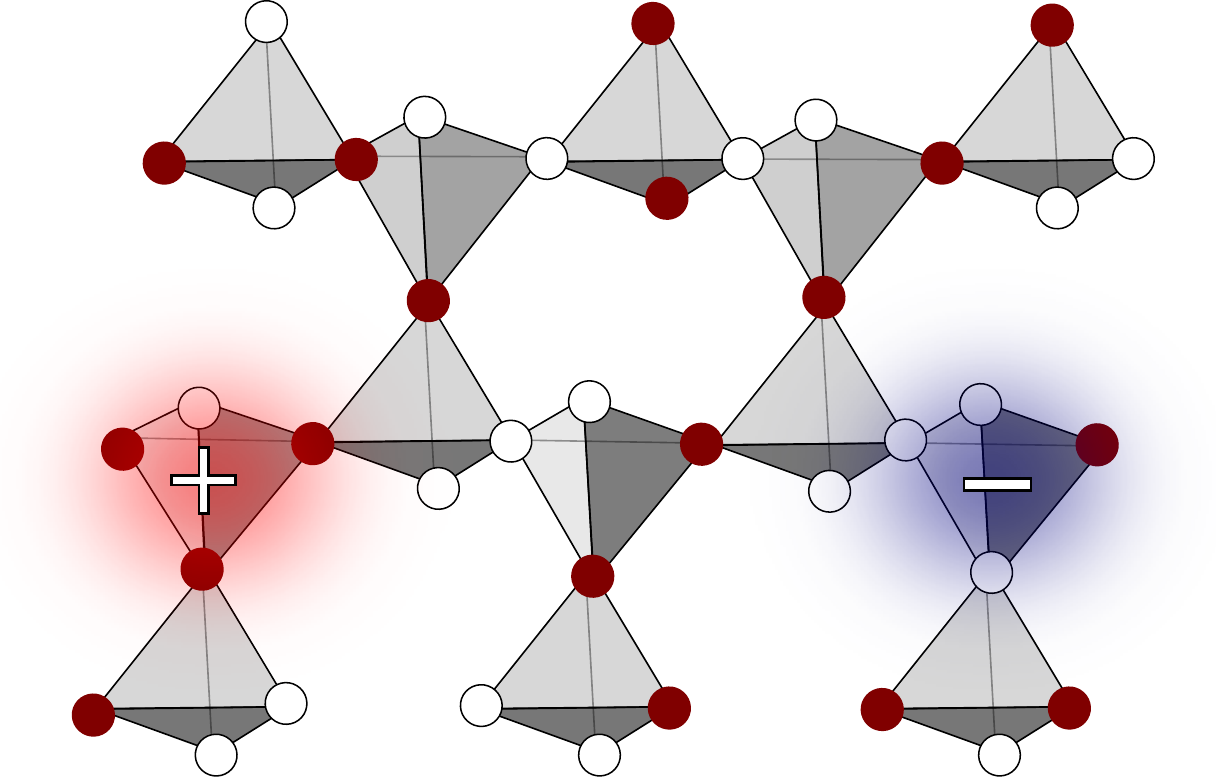}
\label{fig:charge_coulomb}
}
\subfigure[]{
\includegraphics[width=9cm]{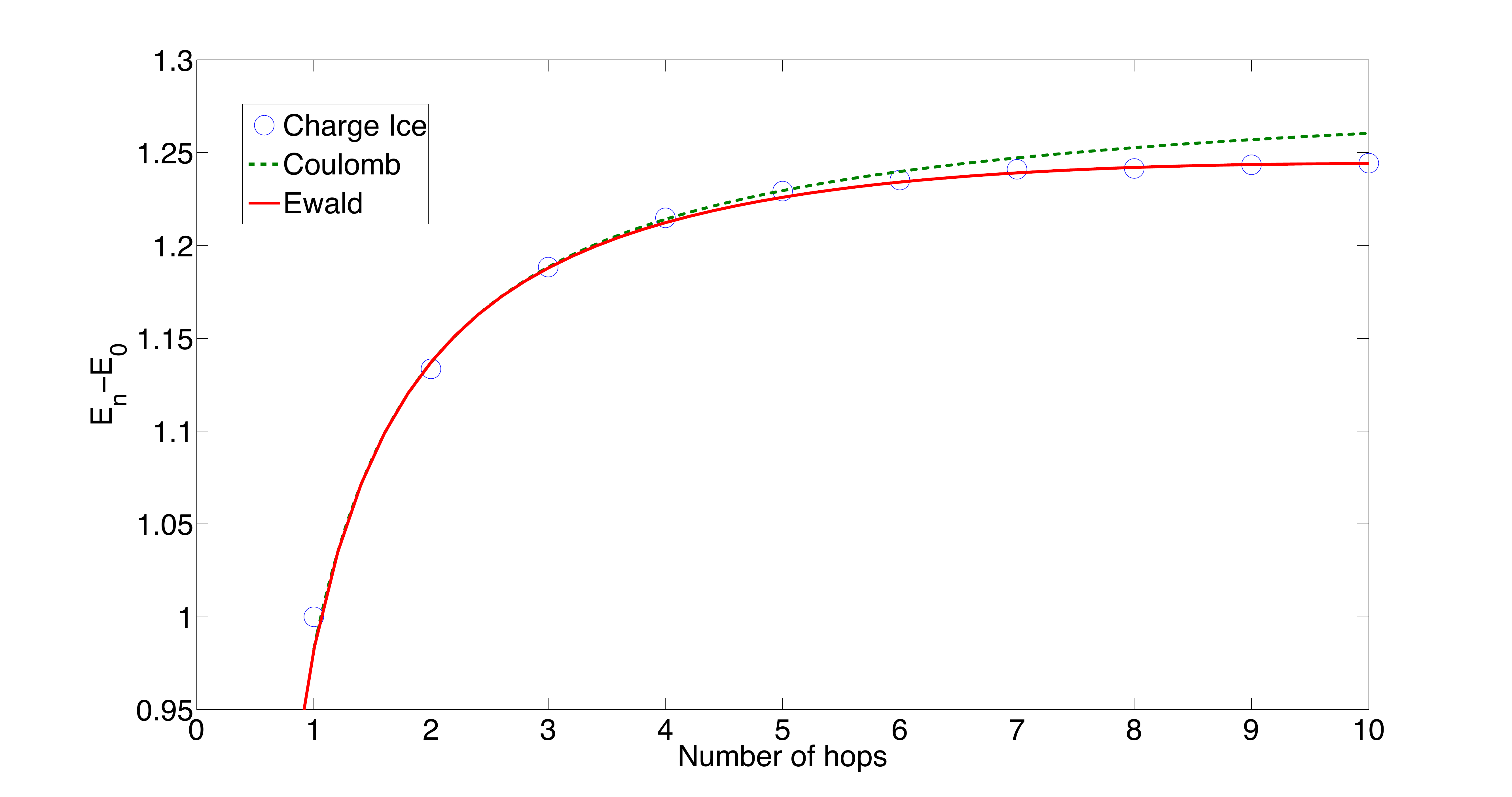}
\label{fig:charge_coulomb}
}
\caption{Panel (a) shows two defects  (fractional charges) of negative and positive charge. Energy as a function of the distance between two opposite charges as a function of distance is shown in panel (b) . The $1/r$ potential provides a good fit when the defect separation is smaller than the length of the system. The Coulomb potential adjusted to the finite periodic cell (the Ewald potential) provides an excellent fit to the data.}
\label{fig:charge_coulomb}
\end{figure}

\section{Results in Applied Electric Field}
\label{sec:EField}

Previously we have seen that the charge ice model exhibits a Coulomb phase characterized by a short-range constraint on the total charge on each tetrahedron which is unstable at lower temperatures to a charge ordered phase. Much of this qualitative physics is shared by dipolar spin ice and, indeed, we have shown that the dipolar spin ice states and the charge ice states are related by a staggering of the dipole moment on half of the tetrahedra. 

In this section, we consider the equilibrium response of charge ice to a uniform static electric field. We will see that it differs markedly from the analogue problem of dipolar spin ice in a magnetic field. We begin with a toy model - nearest neighbor charge ice in an electric field which we will compare with nearest neighbor spin ice in a magnetic field. After this, we consider the Coulombic charge ice model in an electric field.

\subsection{Nearest Neighbor Model}

The model is 
\[  H = V \sum_{\langle i,j\rangle } Q_{i}Q_{j}  - \mathbf{E} \cdot \sum_{i} Q_{i} \mathbf{R}_{i} \]
where we work in units where the elementary `charges' on each site can take the values $Q_{i}=\pm 1$. The total charge on the lattice is zero. 

For definiteness, let us choose the electric field to point in the $[001]$ crystallographic direction (the vertical direction in the panels of Fig.~\ref{FIG:4STATES}). The analysis can be generalized to any field direction with appropriate boundaries. The model is now
\[  H_{E} = V \sum_{\langle i,j\rangle } Q_{i}Q_{j} - E \sum_{i} Q_{i} Z_{i}   \]
where $Z_i$ is the coordinate of site $i$ in the field direction. Evidently, the constancy of the net charge ensures that the energy is independent of the origin of $Z$. We recall that the pyrochlore lattice is a network of corner-sharing tetrahedra. Each vertex is shared by two tetrahedra. Suppose the centre of tetrahedron $T$ is at $Z_{T}$ and that site $i$ is shared by tetrahedra $T_A$ and $T_B$. Then
\[  -E Q_{i}Z_{i} = -\frac{E}{2} Q_{i} (Z_{T_{A}} + b_{i}) -\frac{E}{2} Q_{i} (Z_{T_{B}} - b_{i})      \]
where $b_i = 1/2$ is the location in the $z$ direction of site $i$ within a tetrahedron.
Let $Q_T \equiv \sum_{i\in T} Q_{i}$. Up to a constant shift in the energy,
\[  H_{E} = \frac{1}{2} \sum_{T} \left( Q_{T} - \frac{E}{2} Z_{T} \right)^{2}   -E\sum_{i \in {\rm surface}} Q_{i}Z_{i}.    \]
Hence the total energy is a sum over positive semidefinite terms on each tetrahedron and a term that depends only on the charge configuration on the surfaces perpendicular to the electric field. We cannot neglect the latter terms - although they are surface terms they make an extensive contribution to the energy.

We consider the surface term first. When the electric field is zero, the minimum energy solution is the one with zero net charge on each tetrahedron as we expect. In nonzero field, the surface terms favor a net charge at the surface for arbitrarily small fields. The ground state is therefore one with a net polarization obtained by transferring charge from one surface to the opposite surface. In order to preserve the ice rule, the charges are arranged such that the charge are opposite on adjacent layers. The requirement that the charge be opposite on opposite faces is consistent with the ice rule only when there is an even number of layers. In odd layered systems, owing to the geometrical obstruction a net polarization would involve introducing a defect on a layer where the ice rule is not satisfied. Since this costs a subextensive amount of energy, the even-odd distinction disappears in the limit of large enough systems. We have checked this using Monte Carlo simulations at finite temperature.

\begin{figure}[h!]
\begin{center}
\includegraphics[width=8cm]{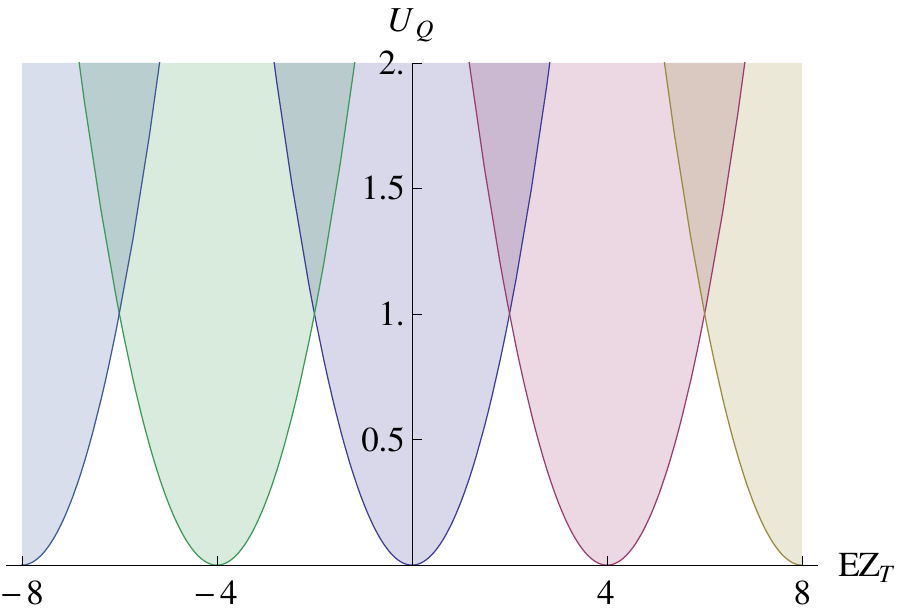}
\end{center}
\caption{Figure showing $U_{Q}=(Q_{T}-(EZ_{T}/2))^2$ for the possible tetrahedron charges $0$, $\pm 2$ and $\pm 4$. The minimum energy may live on any of these five branches depending on the location of the tetrahedron in the sample and on the magnitude of the electric field. \label{fig:soln}}
\end{figure}

For the bulk terms, we minimize 
\[  \left( Q_{T} - \frac{E}{2} Z_{T} \right)^2  \]
subject to overall charge neutrality noting that the charge on a tetrahedron can be $0$, $\pm 2$ or $\pm 4$. It is convenient to place the origin of the coordinate system at the centre of the sample to make the symmetry $Z\rightarrow -Z$ transparent. Figure~\ref{fig:soln} shows this function plotted for the possible tetrahedral charges as $Z_T$ is varied. The minimum energy state of a given tetrahedron can be found by tracing out the minimum of these functions. It follows that, for a slab of thickness $L<(2/E)$, the bulk terms are minimized by having zero charge on all tetrahedra. For thicker samples, the bulk terms favor a net polarization across the sample. By inspection of the energy terms, one can see that for a sample of fixed thickness $L\gg a$ where $a$ is the interlayer distance, an increasing electric field causes the sample to polarize from the edge. This observation is borne out by Monte Carlo simulations.

The nearest neighbor spin ice model is strongly anisotropic which is reflected in its magnetic field response. For example, in a weak $[001]$ magnetic field, the spins polarize along the field as far as possible given the Ising constraint. In a $[111]$ field, there is a magnetization plateau as the $[111]$ Ising spin is pinned along the field, leaving an extensive degeneracy in the kagome planes perpendicular to the field. These features are shared by the long-range interacting dipolar spin ice model and by the spin ice materials. 

The nearest neighbor model discussed in this section is, by comparison, isotropic in the sense that a sufficiently large slab undergoes charge separation even in a small electric field. For thin slabs and small fields, the response in an electric field is comparable to that of spin ice including the presence of kagome ice. In the next section, we consider the full long-range interacting charge model.

\subsection{Coulombic Charge Ice}

The discussion of the previous section concentrated on the case where a uniform electric field is applied to an ice model but where we have made the unphysical simplification that the charges cannot screen the electric field. In the charge ordered phase of the long range interacting model, the charge gap forbids any nontrivial response to an electric field below some threshold field controlled by the size of the gap. However, the effect of screening is crucial to the behavior of the Coulomb phase. 

Monte Carlo simulations of the Coulombic charge ice model were performed employing the three dimensional EW3DLC Ewald summation method appropriate to the slab geometry which carries out an anisotropic sum over periodic images reflecting the slab geometry as well as placing empty space between image slabs and suppressing the potential between them. \cite{Yeh1999} We apply an electric field in the $[001]$ direction. In contrast to the nearest neighbor model, the long range model does not enter an ice state with homogeneous polarization in an electric field. Instead, we find that the simulation cells polarize in an electric field with the charge accumulating at the edge. More precisely, the averaged charge per layer decays into the bulk with some correlation length. This correlation length grows upon cooling into the Coulomb phase. We have measured the thermal averaged polarization density $\rho_{\rm P}$ as a function of electric field where
\[   \rho_{\rm P} \equiv \frac{1}{V} \sum_i Q_i Z_i,     \] 
$Z_i$ being the direction in which the field is applied. We expect to find that this is proportional to the electric field as expected when the system screens the electric field. Indeed, at high temperatures where the correlation length is small, we find $\rho_{\rm P}\propto E$ while departures from this relation occur within the Coulomb phase for small system sizes and large fields. 

It is interesting to know how this polarization affects the Coulomb phase in the bulk. We know from the case of dipolar spin ice that the application of a magnetic field magnetizes the bulk which involves a release of entropy. Viewed from the point of view of effective charges, the magnetic field creates positively and negatively charged pairs which hop to opposite faces along the magnetic field lines. When a pair of effective charges is separated a so-called Dirac string, which runs between the charges, is extended and this string corresponds to fixed spin directions. So the proliferation of Dirac strings results in a loss of entropy from the bulk. 

\begin{figure}[h!]
\begin{center}
\includegraphics[width=9cm]{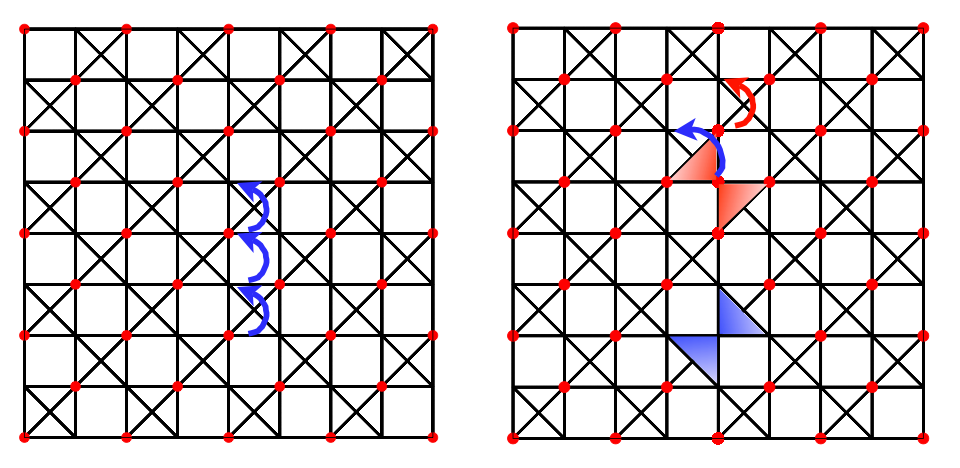}
\end{center}
\caption{Charge ice configuration (left) on a checkerboard lattice - the $[001]$ projection of the pyrochlore. Positive charges only are shown (as red dots). The sequence of hoppings indicated by blue arrows on the left figure produce the charge configuration in the right figure with two pairs of tetrahedral defects. The blue arrow on the right figure shows how to hop the defect pair while the red arrow shows how to hop each tetrahedral defect independently. \label{fig:string}}
\end{figure}

In charge ice, charges can be separated in pairs with no accompanying Dirac string. To see this, refer to Fig.~\ref{fig:string} where we have shown a charge ice configuration on a projection of the pyrochlore lattice onto two dimensions. By hopping a pair of charges as indicated, we can create two pairs of adjacent tetrahedra where the ice rule is violated. The adjacent tetrahedra have the same effective charge. These pairs of defected tetrahedra cannot occur in dipolar spin ice. One can hop these defected tetrahedra together as a single unit with no effect on the background ice configuration. In other words, no Dirac strings are created in the process. Of course, Dirac strings are not forbidden in charge ice; starting from one pair of defected tetrahedra, we can hop a sequence of charges to separate these tetrahedra resulting in the formation of a string between the defects. 

\begin{figure}[h!]
\begin{center}
\includegraphics[width=9cm]{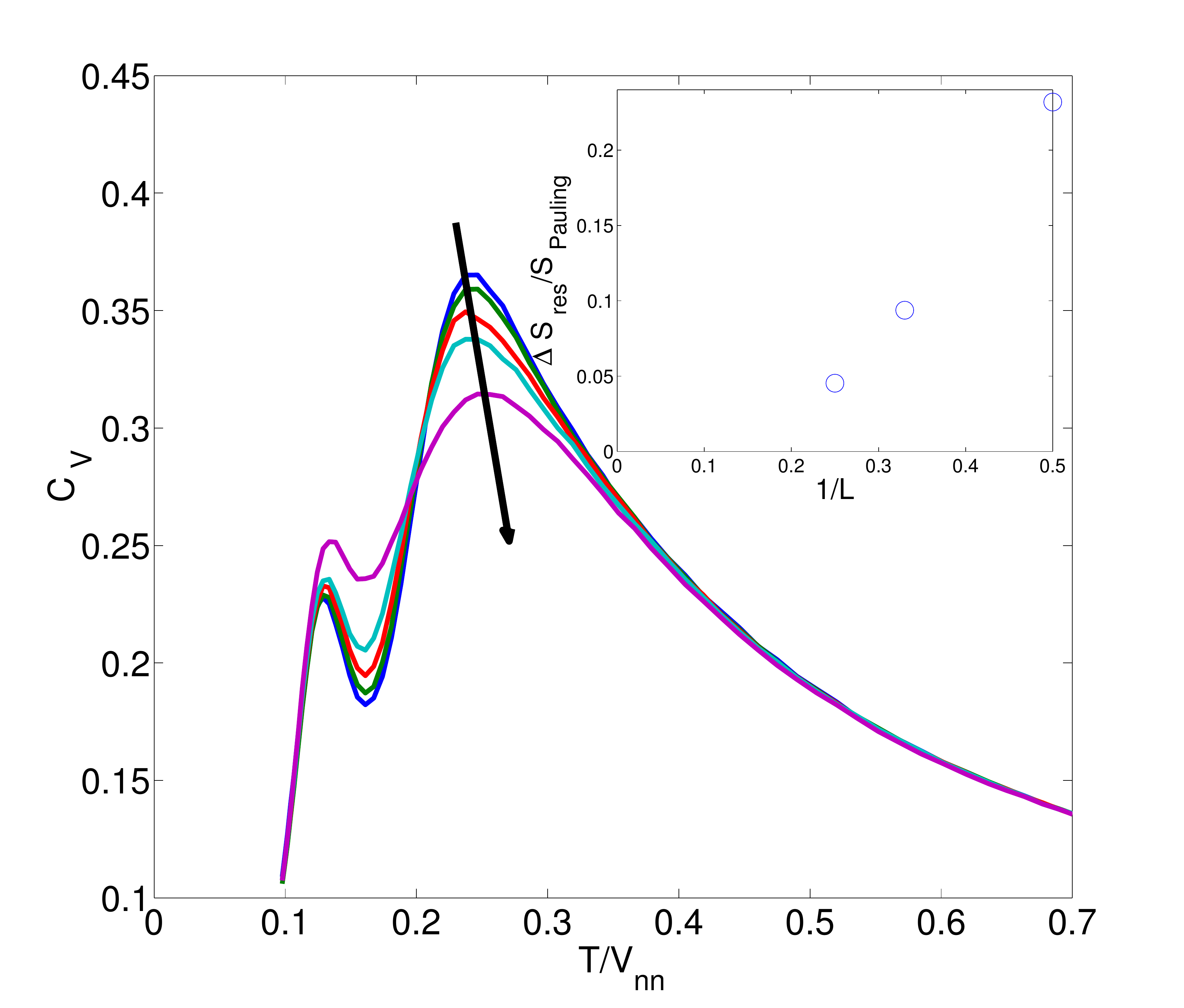}
\end{center}
\caption{Heat capacities for $L=2$ for progressively increasing $[001]$ electric field strengths. The inset is a plot of the gain of residual entropy in a fixed electric field relative to the zero field residual entropy for three different system sizes. \label{fig:Sres}}
\end{figure}

Having seen that the bulk entropy need not be lost although charges are moved to opposite surfaces in an electric field, we turn to the simulation results. We measured the residual entropy for different applied electric fields and system sizes with up to $1024$ lattice sites. The residual entropy $S_{\rm res}$ is measured from the base of the finite size phase transition heat capacity peak. $S_{\rm res}$ systematically increases as the field increases from zero in contrast to the behavior of spin ice in a magnetic field. This behavior arises from the increase in the configurational entropy at the surface of the system when a few extra charges are added on the average. Since we are observing contributions from the surface, we should expect the increase in $S_{\rm res}$ to scale to zero in the thermodynamic limit. Fig.~\ref{fig:Sres} shows the difference in the residual entropy at two different electric fields $\Delta S_{\rm res}= S_{\rm res}(E/V_{\rm nn}=0.18)-S_{\rm res}(E/V_{\rm nn}=0)$ for three system sizes. The tendency is for the change in residual entropy to go to zero as the system size increases though the system sizes are not large enough to see the expected $1/L$ scaling. 

In summary, we have found evidence based on retention of residual entropy that screening in charge ice protects the Coulomb phase in the bulk from polarizing in an applied electric field.  

\section{Remarks on Transport Properties in Charge Ice}
\label{sec:pseudogap}

In this section, we briefly consider the problem of adding a hopping term to the hitherto classical model. When the hopping kinetic energy $t$ exceeds the Coulomb energy, the system is simply a Fermi liquid so we concentrate on the semiclassical case where the hopping is small compared to the Coulomb energy. In this case, there are three distinct temperature regimes: (i) below the charge ordering temperature the transport is expected to be thermally activated leading to a conductivity varying as $\sigma\sim \exp(-U/k_{\rm B}T)$ where $U$ is the charge gap, (ii) the intermediate charge ice regime which extends from $T_c$ to some scale $T^{\star}$ and (iii) high temperatures $T>T^{\star}$ where scattering with phonons leads to $\sigma\sim T^{-1}$. 

\begin{figure}[h!]
\includegraphics[width=8cm]{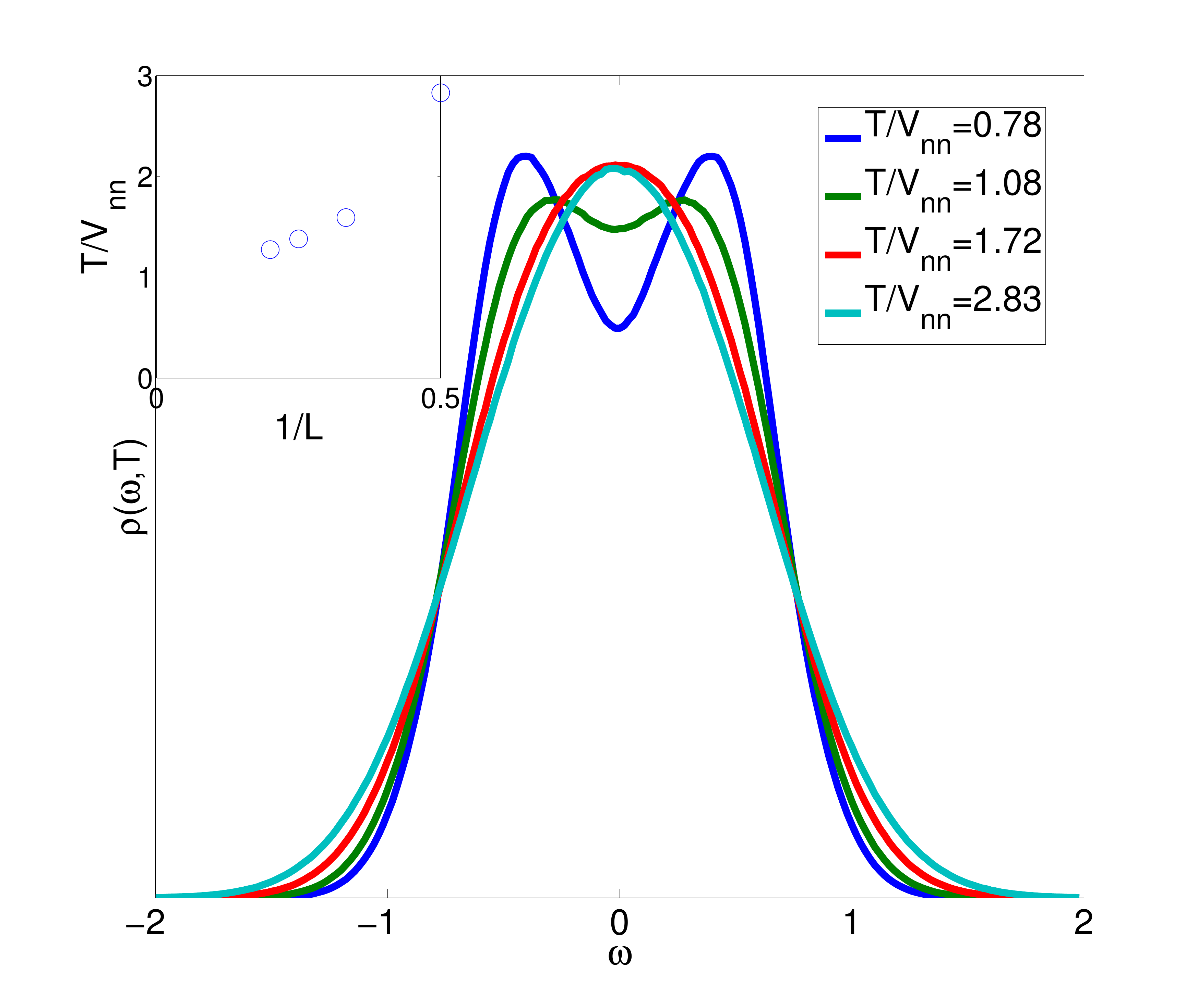}
\caption{Single particle density of states of the charge ice model for $L=4$ in the classical limit of the hopping $t\rightarrow 0$. Four different temperatures are illustrated showing the onset of the pseudogap regime. The inset gives the onset temperature of the double peak structure for three different system sizes.}
\label{fig:DOS}
\end{figure}

In the strongly correlated charge ice regime, we can borrow insights from Ref.~[\onlinecite{Pramudya2011}] in which the problem of particles interacting through a $V(R)\sim R^{-\alpha}$ potential was considered. By computing the conductivity within EDMFT for this problem, the authors found an extended regime above the charge ordering transition where the conductivity increases weakly (almost linearly) with increasing temperature. This poor insulating behaviour coincides with the appearance of a pseudogap in the single particle density of states $\rho(\omega)$ characterized by a bimodal distribution with non vanishing spectral weight at $\omega=0$. This pseudogap regime exists between $T_c$ and $T^{\star}$ - the crosssover temperature being weakly dependent on the range of the interaction $\alpha$. In the charge ice model, we expect to find a similar pseudogap within the intermediate temperature regime and corresponding poor insulating transport. To see this, we have measured the single particle density of states which, in the classical limit ($t=0$),  is the distribution of electrostatic potentials
\[   \rho(\omega,T) = \left\langle \sum_i \delta\left(\omega -  \sum_{j} V(r_{ij}) \right)  \right\rangle.  \]
Fig.~\ref{fig:DOS} shows the onset of the pseudogap for $L=4$. 

We obtain a value of $T^{\star}/T_c \approx 10$ which is much larger than the corresponding ratio found for the half-filled cubic lattice $(T^{\star}/T_c)_{\rm cubic} \approx 2.5$. \cite{Pramudya2011} Since the ratio is known to increase as the range of the power law interaction increases \cite{Pramudya2011} it is natural to regard the ratio as a measure of the frustration of the interactions. 

It is worth noting that the conductivity of magnetite increases weakly with temperature above the Verwey transition ($T_c \approx 120$ K) up to about $350$ K whereupon it decreases up to the ferrimagnetic transition at about $850$ K. It appears likely the frustration of the Coulomb interactions leads to the roughly linear dependence of the $T_c <T<350 $K conductivity although the temperature window over which this is observed in magnetite is much smaller than expected in the charge ice model. It would be interesting to follow this up with a more direct probe of the charge correlations in magnetite above the Verwey transition.

\section{Discussion}
\label{sec:discussion}

Early progress in the theory of the mixed valence material magnetite led to the proposal that the Verwey transition in that material might be a charge ordering transition that could be captured by a model of charges on a pyrochlore lattice with half of the sites occupied by one of the two species of iron ions. While the transition in magnetite appears to be associated with the onset of charge order, the charge order is not that predicted by the original charge model. This and other complications including a structural transition in magnetite have apparently made the charge ice model obsolete. However, in the intervening years, insights into the phenomenology of geometrically frustration have become available and have made it timely to explore the implications of the frustration of Coulomb interactions on the pyrochlore lattice. This article adds a new perspective to the physics of Coulomb phases which have been explored in detail in the spin ice materials but which may be observed in wide range of different systems including charge \cite{Pauling1935,Henley2011}, magnetic \cite{Bramwell01} and orbital \cite{Chern2011} degrees of freedom.

Using classical Monte Carlo on the Coulombic charge ice model we find a temperature window where there is an ice regime as indicated by a Pauling residual entropy. At lower temperatures, simulations fall out of equilibrium without the assistance of nonlocal Monte Carlo moves. Such nonlocal moves allow one to access the low temperature first-order transition to a charge ordered state. A study of the spectrum of interactions reveals the existence of low energy ice states and the weak lifting of degeneracy of the lowest ice bands compatible with the long-range order observed using Monte Carlo. We show that the long wavelength corrections to dipolar correlations coming from the degeneracy breaking correspond to a relatively short range $r^{-5}$ interaction. Raising the temperature from the ice regime produces point-like excitations carrying a fractional charge.

There are strong parallels between the above properties of Coulombic charge ice and those of dipolar spin ice. In fact, these two long-range interacting models are closely related. The dumbbell picture in which pyrochlore Ising spins are separated into dumbbells with equal and opposite charges at their ends has an analogue for the charge ice model. Then both ice models can be thought of as descended from a model of dipoles sitting on a diamond lattice. The long-range ordered structure of dipolar spin ice is the unconstrained ground state of this diamond lattice model. 

Turning to the experimental observables of charge ice, we have computed the charge-charge correlations in the ice regime which exhibit smeared pinch points just above the charge ordering temperature. In a static electric field, charge ice polarises with charge density appearing at the surfaces, screening the electric field within so the ice state is preserved by weak electric fields. This is in sharp contrast to spin ice which immediately loses entropy in an applied magnetic field. From the point of view of the elementary excitations, charge ice differs from spin ice in being able to heal Dirac strings that form when fractional charges are separated. Finally, we have considered the transport properties in the ice regime noting that the single particle density of states becomes bimodal. The resulting pseudogap has been associated \cite{Pramudya2011} with poor conducting behavior that is characteristic of materials above a charge ordering transition - in particular in magnetite. 

Perhaps the most interesting open questions arising from this work center on the relationship of the charge ice model to mixed valence materials including magnetite and its relatives where the Coulomb interaction may be highly frustrated. One possible step in this direction is to look for signatures of pinch point features in charge correlations using resonant X-ray scattering. Also, it would also be interesting to study the pseudogap picture of charge transport within a microscopic model for magnetite. In these ways and perhaps others, one might hope to shed further light on the, often formidably, rich phenomenology of the mixed valence spinels. 

\section*{Acknowledgment}
We gratefully acknowledge Roderich Moessner for various important insights during the development of this work.
\appendix

\section{Energetics of small clusters}
\label{appx:1}

\begin{figure}[h!]
\includegraphics[width=9cm]{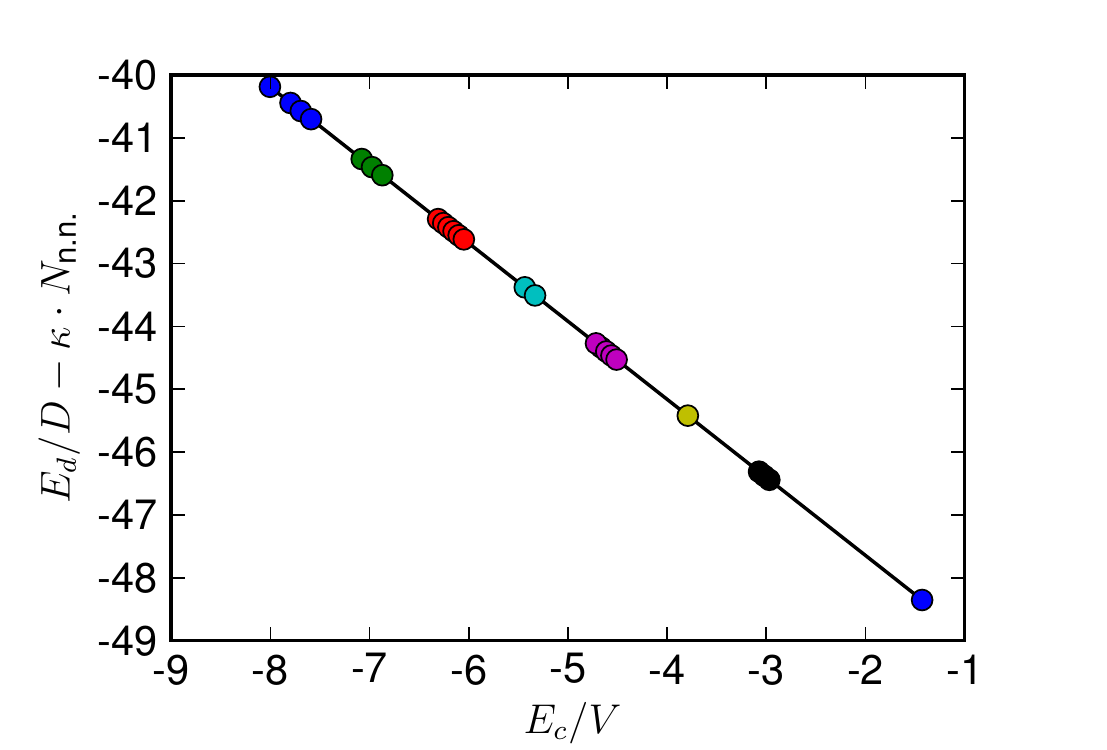}
\caption{ Spectra of the dipolar spin ($E_d$) and charge ice model ($E_c$) for states commensurate with a $16$ site cubic unit cell are identical up to a sign change for all sectors with different number of defects of the ice rule. The constant used to align all sectors is approximately $\kappa \approx 3.71$. \label{FIG:COMPARE16} } 
\end{figure}

In this appendix we consider the spectrum of states for the dipolar and charge ice models on the smallest cubic unit cell which has $16$ sites. In particular, we report a simple pattern in the energies of the states on this small cluster. Because of the small unit cell size, we are able to enumerate all possible states. There are in total $12\ 870$ different charge (spin) configurations with no net charge which correspond to 25 inequivalent states (Other states can be obtained using symmetry operations). Only 90 configurations fulfil the ice rule which group into the four inequivalent states shown in Fig.~\ref{FIG:4STATES}. The energies of the two models taking into account  long range dipolar and Coulomb interactions are compared in Fig.~\ref{FIG:COMPARE16}. We find that the spectra of all states are identical up to a sign change and a shift which is proportional to the number of of defects of the tetrahedron rule. The ground state of the Coulombic charge ice is 6-fold degenerate  (Fig~\ref{FIG:4STATES}~a) and the one of spin ice is 12-fold degenerate (Fig~\ref{FIG:4STATES}~d). 

In section~\ref{sec:energetics}, we saw that dipolar ice configurations may be mapped into charge ice configurations. We also saw that, within the dumbbell model on a diamond lattice, charge ice configurations differing in the sign of the dipolar moment on the B tetrahedra map into one another. Let us consider the implications of this fact for the spectra of the two model. Since, we have mapped the charge and dipolar models onto diamond lattice models with some multipole moments on each diamond vertex the energy can be written as a multipole expansion. To lowest order, the energy is simply the total energy of all tetrahedra. 
\[  E^{(0)} =  \frac{1}{2}\sum_{\rm t} \sum_{i(t),j(t),i(t)\neq j(t)}  \frac{q_{i(t)}q_{j(t)}}{r_{i_{i}j(t)}}   \]
where $i(t)$ and $j(t)$ run from $1$ to $4$. For the ice states, all $N/2$ tetrahedra have the same energy. The next terms in the expansion involve interactions between tetrahedra. Now suppose, the interactions only act between nearest neighbour tetrahedra. Then we arrive at our first main result: the terms involving interacting tetrahedra are identical in the charge and dipolar model except for an overall sign change. The inversion of the spectra is clearly exact in this limit.

Since ice states have tetrahedra with zero charge, the leading order interactions are the dipole-dipole interactions. Let us consider the effect of increasing the range of the dipole-dipole interactions in the ice configurations. The $90$ ice states in the cluster are grouped into $4$ distinct energies (shown in Fig~\ref{FIG:4STATES}~a). We have computed, in addition, the interactions between quadrupoles and hence the total energy to this order in the multipole expansion. If we translate the energy of the six-fold degenerate state to zero, normalize the largest energy to unity and reverse the sign of the dipolar model spectrum, we find that the relative energies to this order are $0$, $1/2$, $3/4$ and $1$ reflecting the pattern in the exact energies.

This simple pattern arises from two features of the small cluster. The first is that long range interactions on this small periodic cluster have three inequivalent couplings between pairs of sites: the nearest neighbor, second neighbor and third neighbor. This is a purely geometrical fact. Within a given defect sector (those configurations with a given number and effective charge of tetrahedral defects), the configurations are all degenerate with respect to the nearest neighbor interaction. So, only the second and third neighbor couplings lift the degeneracy meaning that, for each defect sector, we need only consider
\[  \Delta U = \alpha_2 n_2 + \alpha_3 n_3  \] 
where
\[  n_2 = \sum_{\langle \langle i,j \rangle\rangle } q_i q_j    \]
and
\[  n_3 = \sum_{\langle \langle  \langle i,j \rangle\rangle\rangle } q_i q_j    \]
and $\alpha_2$ and $\alpha_3$ are the Ewald coefficients for second and third neighbor interactions respectively. The second important feature of this small cluster is that the degeneracy within a given defect sector can be explored by making a worm move (i.e moving charges along a closed loop of alternating charges) and that, if the worm move increases/decreases $n_2$ by $m$, then $n_3$ decreases/increases by $m$. 

We conclude, that the relative energies of configurations $c$ within a given defect sector are just
\[   \Delta U_c = {\rm const.} + p_c(\alpha_2 - \alpha_3)     \]
where $p_c$ is an integer. It turns out that $\alpha_2 - \alpha_3$ has an opposite sign for the charge ice and dipolar spin ice models which accounts for the flipping of the spectra within the distinct defect sectors.

We note that the previous argument relies on the special properties of the $16$ site cluster. There is no reason to expect that the simple relationship between the charge and dipolar spin ice spectra persists to larger clusters and, indeed, it does not. Recall, however, from Section~\ref{sec:energetics} that the staggering of dipoles on diamond $B$ sites relates the charge ice and dipolar spin ice states. Since, to leading order, the degeneracy of the ice states is broken by the interaction between the moments on the diamond sites, we expect that the spectra of ice states in the dipolar spin ice and charge ice models has some weak anti correlation for arbitrary system sizes.

\section{Dumbbell Model}
\label{appx:2}

In this appendix, we present an analysis of the dumbbell models for charge ice and dipolar spin ice. The question we wish to answer is whether the dumbbell models faithfully capture the mean field behavior of dipolar spin ice and charge ice. Recall that the dumbbell model is obtained from dipolar spin ice by replacing each Ising spin by a dipole of equal and opposite charges while for the charge model we break each charge into a pair of equal charges. Each pair of charges is connected by a line oriented in the local $\langle 111\rangle$ directions. The idea is illustrated in Fig.~\ref{fig:dumbbell}. 

We can formulate the dumbbell models equivalently as a diamond lattice model with a tetrahedral basis and charges placed on the resulting lattice sites. This procedure doubles the number of degrees of freedom compared to the original models. Also, whereas the original charge and dipolar ice models are defined only up to an overall scale, the dumbbell model has a tuning parameter, which we call $\epsilon$, which tunes the edge length of the tetrahedral basis between the pyrochlore limit $\epsilon=0$ and the diamond lattice limit $\epsilon=1$. We obtain the spectrum of interactions for this model as was done for charge ice in Section~\ref{sec:spectrum}. The Bravais lattice is fcc with a two site basis for the diamond sites and, for each diamond site, a tetrahedron of sites. This gives an eight charge basis in total and hence eight bands in reciprocal space.

We then impose a constraint on the charges of the dumbbell model to separate out the charge ice bands from the dipolar ice bands. One way of doing this is to carry out a projection in the following way. First carry out a unitary transformation $\mathbf{D}$ to obtain the symmetric and antisymmetric combinations of the two charges that form the dumbbells; there are four such pairs of charges. Starting from the $8\times 8$ interaction matrix $\mathbf{V}(\mathbf{q})$ for the pair of charges, we obtain,
\[  \mathbf{D}  \mathbf{V}(\mathbf{q}) \mathbf{D}^{\dagger}.   \]
Then, to obtain the symmetric interactions and the antisymmetric interactions, we project out the upper left $4\times 4$ block and the lower right $4\times 4$ block respectively. Another way of enforcing the charge constraint is to add interactions of the form
\[  \sum_{D\in \left\{ \rm Dumbbells\right\}} \left( Q_{D,1} \pm Q_{D,2}   \right)^{2}   \]
where charge $Q_{D,1}$ is charge $1$ belonging to dumbbell $D$. This penalizes dumbbells that do not have equal and opposite (plus sign, dipolar) or equal (minus sign, charge) configurations. These two ways of obtaining the dipole and charge ice bands give qualitatively identical results.

Computing the eigenvalues of the constrained dumbbell model for both the charge and dipolar cases, one may determine the ground states of the two models. For the charge case the minimum eigenvalue is at $\mathbf{q}=0$ and for the dipole model the minimum is at $\mathbf{q}=(0,0,2\pi)$. These are respectively the mean field ordering wavevectors of the two models. The ordering wavevector does not depend on the value of $\epsilon$. These results are consistent with the interaction spectra of the original models and with Monte Carlo simulations. In addition, we find that (i) the two lowest bands have little dispersion compared to the next highest pair of bands and (ii) there is no band crossing as a function of $\epsilon$. Taken together, these facts reinforce the idea that the dumbbell picture captures the physics of the original models faithfully.

\begin{figure}
\subfigure[Charge Model]{
\includegraphics[width=9cm,clip]{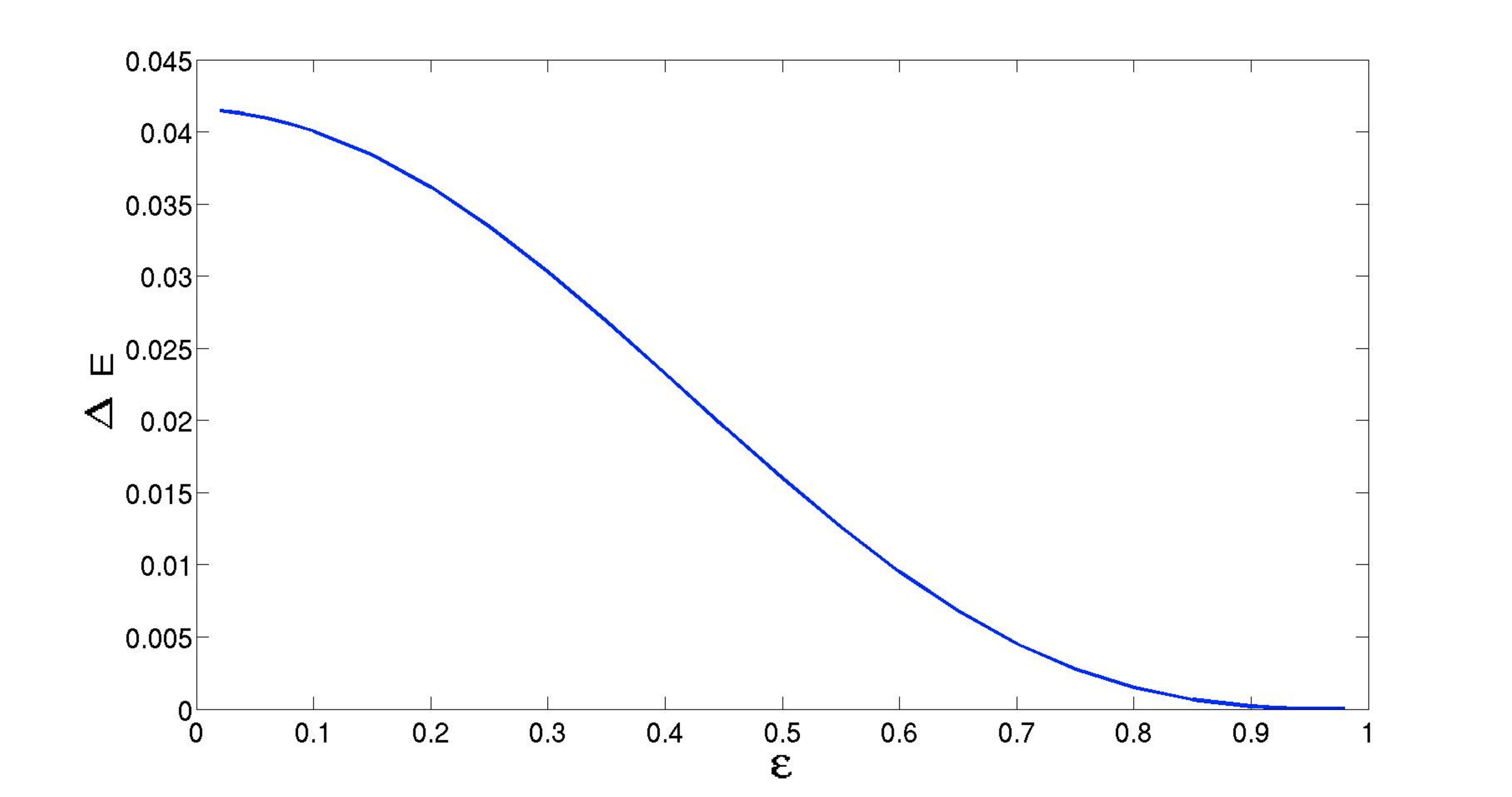}
}
\subfigure[Dipolar Model]{
\includegraphics[width=9cm,clip]{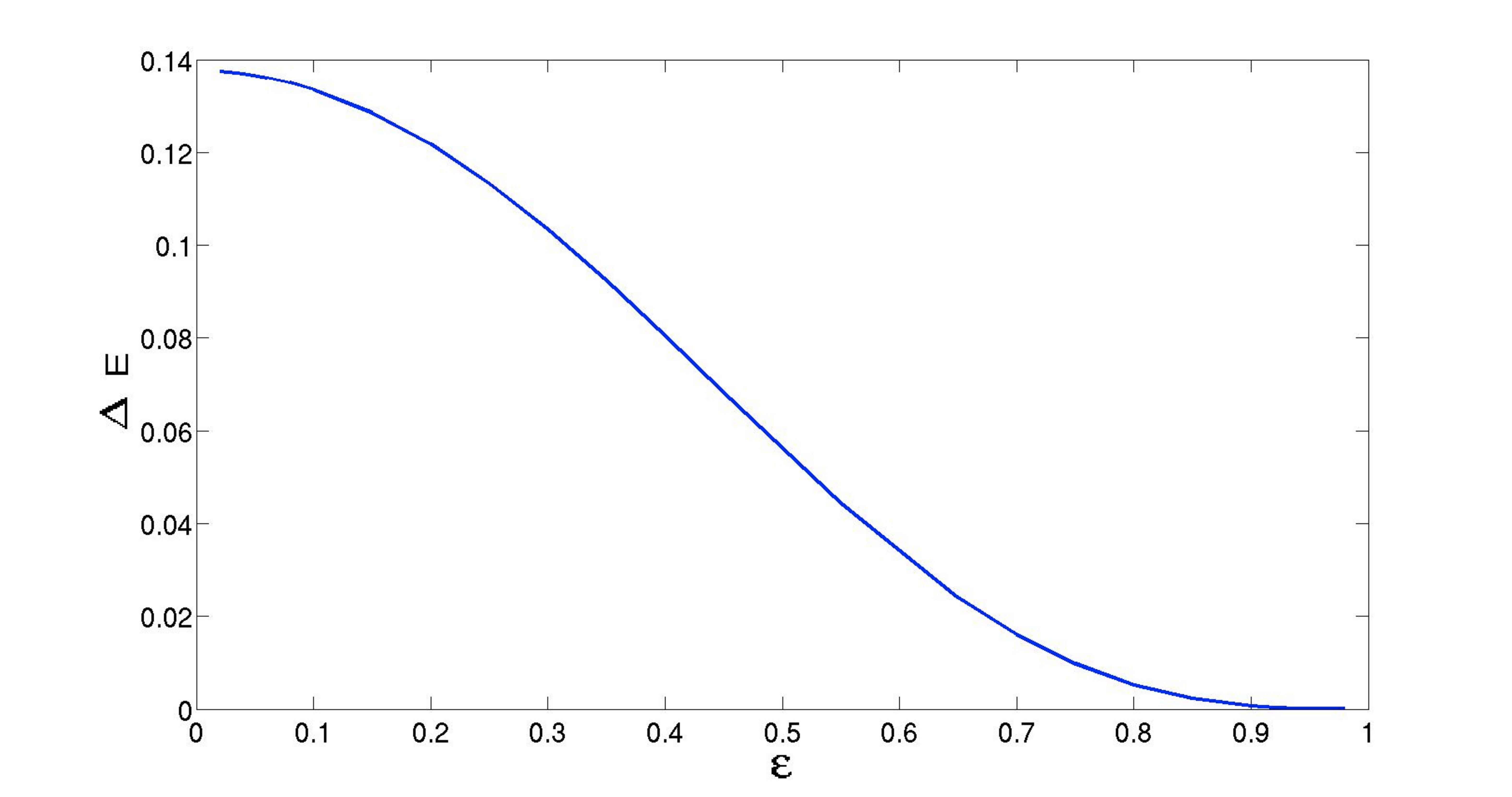}
}
\caption{\label{fig:dumbbell_bandwidth} Figure showing the variation in the bandwidth of the two lowest bands in the constrained dumbbell model.}
\end{figure}

Fig.~\ref{fig:dumbbell_bandwidth} shows the variation in the total bandwidth of the two lowest bands as a function of $\epsilon$ normalized to the total bandwidth of the next highest band. In the dipolar case, the $\epsilon=0$ limit (for which the charges come together on tetrahedral sites) is finite owing to the normalization. The bandwidths for both the charge and dipolar models tend to zero as $\epsilon\rightarrow 1$ because the ice rule is satisfied. The variation in the bandwidth close to the limiting value $\epsilon=1$ is a power law with exponent $2$. The normalized bandwidth is much smaller after imposing the charge model constraint than in the dipolar model. However, the absolute value of the charge model bandwidth is larger than that of the dipolar model after imposing the further constraint that the dipole moment should not change with $\epsilon$.

\section{Overlap between Flat Band States}
\label{appx:3}

We have studied the overlap between eigenstates of the interactions for different pyrochlore ice models. The three models we consider are the nearest neighbor spin ice model, the Coulombic charge ice model and dipolar spin ice. In all three cases, of the four bands in the spectrum of interactions, the lowest two bands are either completely flat (nearest neighbor model) or almost so compared to the bandwidth of the two higher energy bands. Let $\vert u_{\mathbf{q},{\rm NN}}^{a}\rangle$ and $\vert u_{\mathbf{q},{\rm CCI}}^{a}\rangle$ be respectively the eigenstates for the nearest neighbor model and the Coulombic charge ice model in the $a$th band with $a=1,2$ being the flat bands. Then we compute the overlap $\vert \langle u_{\mathbf{q},{\rm NN}}^{a} \vert u_{\mathbf{q},{\rm CCI}}^{b} \rangle \vert$ over some region of reciprocal space. We find, after taking suitable linear combinations of the flat band eigenstates of the nearest neighbor model maintaining the orthonormality of the pair of flat band states,  that the eigenstates of the two models are roughly in one-to-one correspondence: $\vert \langle u_{\mathbf{q},{\rm NN}}^{a} \vert u_{\mathbf{q},{\rm CCI}}^{b} \rangle \vert \approx \delta^{ab}$. There is a small degree of overlap between the flat bands of the nearest neighbor model and the excited bands of the charge ice model that varies as a function of $\mathbf{q}$ and, at maximum, is of the order of $10 \%$. A similar result holds when we consider the eigenstates of the dipolar spin ice model $\vert u_{\mathbf{q},{\rm DSI}}^{a}\rangle$. We find the overlap $\mathcal{O}_{\rm NN,DSI}\equiv \vert \langle u_{\mathbf{q},{\rm DSI}}^{a} \vert u_{\mathbf{q},{\rm CCI}}^{b} \rangle \vert$. Here, the overlap between the two lowest bands is complicated slightly by band crossings. These band crossings result in discontinuous changes in the overlap between $\mathcal{O}_{\rm NN,DSI}\approx 0$ and $\mathcal{O}_{\rm NN,DSI}\approx 1$. Aside from the band crossings the eigenstates of the charge and dipolar models are in close correspondence.

\bibliography{corr}

\end{document}